\documentclass[useAMS, subeqn, usenatbib]{mn2e}
\usepackage{epsf,times}
\usepackage{natbib}
\usepackage{graphicx}
\usepackage{epsfig}
\usepackage{hyperref}
\usepackage{comment}
\usepackage{multirow}
\usepackage{amsmath}
\usepackage{amssymb}
\usepackage{wasysym}
\usepackage{pdflscape}
\usepackage{color}
\usepackage{pifont}

\def\mnras{MNRAS}
\def\aj{AJ}
\def\aap{A\&A}
\def\apj{ApJ}
\def\apjl{ApJ}
\def\pasp{PASP}
\def\apjs{ApJ}

\title[UGPS High PM Sources]{High Proper Motion Objects from the UKIDSS Galactic Plane Survey}
\date{June, 2014}
\author[L. Smith et al.]{Leigh Smith$^1$\thanks{E-mail: L.Smith10@herts.ac.uk}, P.W. Lucas$^1$, R. Bunce$^1$, B. Burningham$^1$, H.R.A. Jones$^1$,
\newauthor 
R.L. Smart$^2$, N. Skrzypek$^3$, D.R. Rodriguez$^4$, J. Faherty$^{5,6}$, G. Barentsen$^1$, 
\newauthor 
J.E. Drew$^1$, A.H. Andrei$^{1,7,8}$, S. Catal{\'a}n$^9$, D.J. Pinfield$^1$, D. Redburn$^1$\\
$^1$ Centre for Astrophysics Research, Science and Technology Research Institute, University of Hertfordshire, Hatfield AL10 9AB, UK\\
$^2$ Istituto Nazionale di Astrofisica, Osservatorio Astrofisico di Torino, Strada Osservatorio 20, 10025 Pino Torinese, Italy \\
$^3$ Astrophysics Group, Imperial College London, Blackett Laboratory, Prince Consort Road, London SW7 2AZ, UK\\
$^4$ Departamento de Astronomia, Universidad de Chile, Casilla 36-D, Correo Central, Santiago, Chile\\
$^5$ Department of Terrestrial Magnetism, Carnegie Institution of Washington, Washington, DC 20015, USA\\
$^6$ Hubble Fellow\\
$^7$ Observat\'orio Nacional/MCTI, R. General Jos\'e Cristino 77, CEP 20921-400 Rio de Janeiro - RJ, Brazil\\
$^8$ Observat\'orio do Valongo/UFRJ, Ladeira do Pedro Ant\^onio 43, CEP 20080-090 Rio de Janeiro - RJ, Brazil\\
$^9$ Department of Physics, University of Warwick, Coventry, CV4 7AL
}

\begin{document}

\maketitle

\begin{abstract}
 The UKIDSS Galactic Plane Survey (GPS) began in 2005 as a 7 year effort to survey $\sim$1800 deg$^2$ of the northern Galactic plane in the J, H, and K passbands. The survey included a second epoch of K band data, with a baseline of 2 to 8 years, for the purpose of investigating variability and measuring proper motions. We have calculated proper motions for 167 Million sources in a 900~deg$^2$ area located at $l>$ 60$^{\circ}$ in order to search for new high proper motion objects. Visual inspection has verified 617 high proper motion sources ($>$ 200 $mas~yr^{-1}$) down to K$=$17, of which 153 are new discoveries. Among these we have a new spectroscopically confirmed T5 dwarf, an additional T dwarf with estimated type T6, 13 new L dwarf candidates, and two new common proper motion systems containing ultracool dwarf candidates. We provide improved proper motions for an additional 12 high proper motion stars that were independently discovered in the WISE dataset during the course of this investigation.
\end{abstract}


\section{Introduction}
	Source confusion in the Galactic plane reduces the completeness of searches for nearby stars and brown dwarfs and high proper motion sources in general. The two epochs of high resolution UKIDSS GPS data (\citealt{lawrence07}, \citealt{lucas08}) provide a new resource to search for previously missed high proper motion objects, especially brown dwarfs which would typically have been undetected in previous optical searches. It also allows for identification of new high amplitude infrared variable stars \citep{contreras14}.\\
	The thick veil of dust in the Galactic plane is something of a benefit in identifying nearby high proper motion objects in optical surveys since it obscures more distant stars and lessens the problem of source confusion (see \citeauthor{boyd11a} 2011a, Figure 6; \citeauthor{boyd11b} 2011b, Figure 7). However, many of the nearest objects are relatively faint at optical wavelengths and we must turn to the near infrared where they are brighter. Extinction is less of a problem in the near infrared, which leads us back to a greater problem of source confusion. This has allowed many nearby objects to go unidentified until very recently.\\
	\citet{looper07} reported 11 T dwarf discoveries in the 2MASS dataset \citep{skrutskie06}, three of which were found in a search for mid-late T dwarfs at low Galactic latitudes in the 2MASS Point Source Catalogue.
	\citet{phanbao08} detected 26 new ultracool dwarfs (UCDs, generally regarded as M7 and later) in a photometric and proper motion search at low Galactic latitudes in the DEep Near-Infrared Survey of the Southern sky (DENIS, \citealt{epchtein97}). \citet{lucas10} photometrically identified a very cool T dwarf in the UKIDSS Galactic Plane Survey (GPS, \citep{lucas08}).  
	\citet{burningham11b} identified a further two mid-late T dwarfs in the GPS using a similar method.
	\citet{artigau10} identified DENIS~J081730.0-615520, a T6 dwarf at 4.9pc and $b~\sim$ -14$^{\circ}$ in DENIS, as an unmatched source between the DENIS and 2MASS catalogues due to it's high proper motion. 
	\citet{gizis11} and \citet{castro12} identified 2 L dwarfs within 10~pc of the sun at low Galactic latitudes by searching for detections in WISE with no corresponding detection in 2MASS, indicating a high proper motion. \citet{castro13} identified a further 4 L dwarfs by the same method, one of which WISE J040418.01+412735.6 is close to the Galactic plane and a member of the small subclass of unusually red L dwarfs. 
	\citet{folkes12} identified 246 new UCDs with detectable proper motion in their search for UCDs at low Galactic latitudes using the SUPERCOSMOS  and 2MASS surveys. \citet{beamin13} identified an unusually blue L5 dwarf less than 5$^{\circ}$ from the Galactic centre at 17.5~pc distance due to its high proper motion evident in the Vista Variables in the Via Lactea Survey (VVV). \citeauthor{mace13a} (2013a)  and \citet{cushing14} discovered numerous late T dwarfs in the WISE survey, including WISE J192841.35+235604.9 (T6) and WISE~J200050.19+362950.1 (T8) both of which are bright objects in the GPS footprint that are likely to be within 8~pc of the sun (see Section \ref{newTs}).
	\citet{scholz14} used WISE data to identify a 5-7 pc (taking into account the possibility of multiplicity) $\sim$M9 type UCD in the Galactic plane through a photometric selection of candidates followed by identification of those where the nearest 2MASS source was $>$1" from the WISE position. 

	Recently there have been two new all sky proper motion searches using the WISE database by \citet{luhman14a} and \citet{luhman14b}, and \citet{kirkpatrick14}. \citet{luhman13} identified WISE~J104915.57-531906.1, a binary brown dwarf system at 2 $pc$. Given their relative brightness, many of the objects listed above could have been identified in previous surveys but for the effect of source confusion on both colour-based and proper motion-based searches. Given the success of the recent searches with the low resolution 2 epoch WISE dataset, a search for high proper motion stars using the 2 epoch high resolution GPS dataset could be expected to reveal many previously unidentified objects in the solar neighbourhood.
	
	The initial search described here is limited to objects with K~$<$~17~mag, $l~>~60^{\circ}$ and includes only data taken up to March 31st 2013.
	
	This paper is organised as follows. In Section \ref{data} we describe the available data. In Section \ref{method} we briefly describe the proper motion calculation method. In Section \ref{results} we determine the accuracy and reliability of the catalogue. In Section \ref{discoveries} we outline searches undertaken for objects of interest within the catalogue. In Section \ref{summary} we summarise.

\section{Data}\label{data}
\begin{figure}
		\begin{center}
			\begin{tabular}{c}
				\epsfig{file=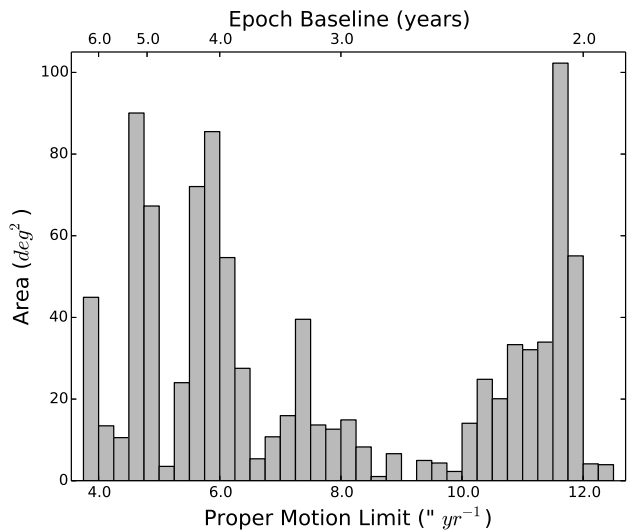,width=1\linewidth,clip=}
			\end{tabular}
			\caption{The area distribution of epoch baselines and maximum proper motion limits of the catalogue.}
			\label{pmlimit}
		\end{center}
	\end{figure}

	The UKIDSS GPS covers 1868 deg$^2$ in J, H, and K passbands to an approximate 5$\sigma$ depth of 18.1 in K. It included a second epoch of K band observations two or more years after the initial epoch. Most of the second epoch K band data are not yet available in the current (8th) GPS data release and have not been processed fully by the WFCAM Science Archive (WSA) team. However, it is retrievable through their Archive Listing tool. Using this we obtained UKIDSS GPS K band FITS file catalogues from observations taken between May 2005 (the start of UKIDSS) and March 31st 2013 and converted them to ASCII format using software provided by the UKIDSS pipeline team \footnote{available at:\\ \url{http://casu.ast.cam.ac.uk/surveys-projects/wfcam/technical}} based at the Cambridge Astronomical Survey Unit. The observations from this date range give us epoch baselines of between 1.9 and 6.4 years (see Figure \ref{pmlimit}). More recent data, from March 31st 2013 to the end of 2013, take the final maximum epoch baseline to approximately 8 years. The catalogues were matched using their telescope pointing positions. In most cases there were two observations per pointing separated by greater than 1.8 years. In these cases we took the earlier observation as the first epoch and the later as the second epoch. In some cases where there were more than two observations per pointing, this is usually due to extra observations that had to be repeated on subsequent nights (e.g. due to low image quality). In such cases we separated those pointing groups into two further groups separated by $>$ 1.8 years between observations. The first epoch K observation would then be the final observation of the earlier group, and the second epoch K observation would be the final observation of the latter group, since once an acceptable observation was made it would not normally be repeated.
	
\begin{figure*}
		\begin{center}
			\begin{tabular}{c}
				\epsfig{file=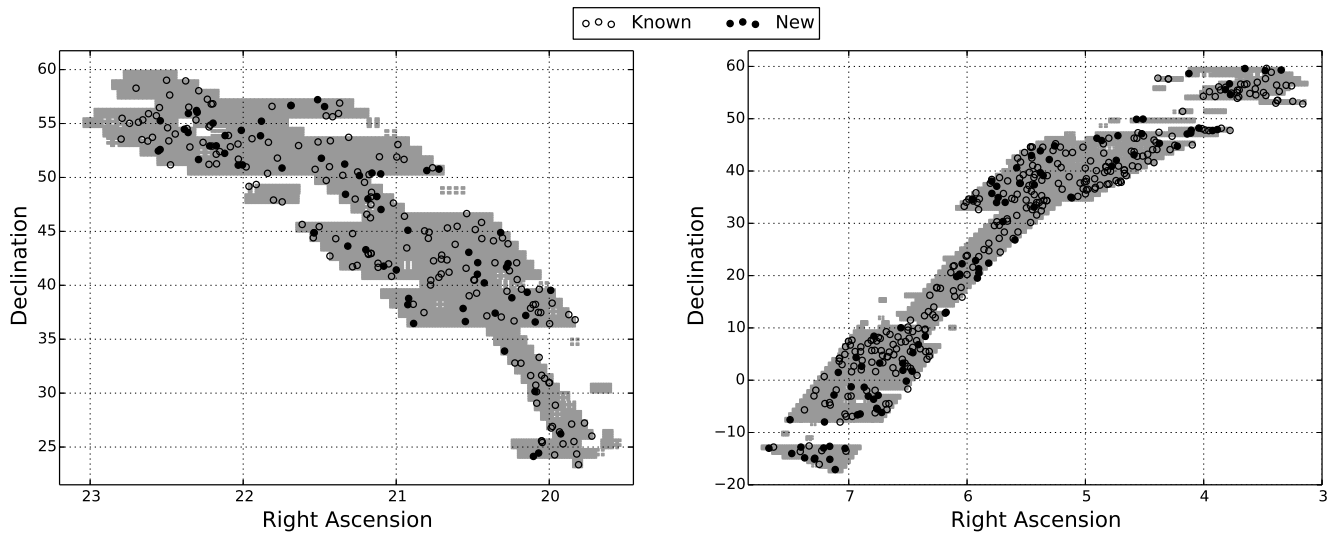,width=1\linewidth,clip=}
			\end{tabular}
			\caption{The coverage of the two epochs of K band data is shown here in grey. Overplotted are the visually verified high proper motion sources. Empty circles are previously identified by other authors, filled circles are new discoveries.}
			\label{coverage}
		\end{center}
	\end{figure*}

	Once we had identified the first and second epoch observations for each pointing we matched sources within the catalogue pairs using a 24" matching radius and an allowance of up to 0.3 K magnitudes difference between sources, returning the best match for each source using a Sky+1d match in STILTS \citep{taylor06}. Following initial tests we chose to limit this search to observations with $l~>$~60$^{\circ}$, in order to keep the number of high proper motion candidates to a manageable level. Since we have a fixed matching radius and epoch baselines ranging from 1.9 to 6.4 years we also have upper proper motion detection limits of between 3.75 and 12.6 "$~yr^{-1}$, see Figure \ref{pmlimit}. The resultant pipeline input catalogue contained $\sim$167 million sources and covered approximately 900 deg$^2$, see Figure \ref{coverage}.
	Note that this method of data collection and initial processing is very similar to that previously used in \citet{smith14} to create a proper motion catalogue for the UKIDSS Large Area Survey. The only differences are an increase in the matching radius and a decrease in the magnitude difference tolerance. The increased matching radius allows us to probe for higher proper motion sources, at the expense of an increase in the rate of mismatches, while not negatively impacting our sensitivity towards low proper motion sources. Since the source density is much greater in the GPS relative to the UKIDSS Large Area Survey (LAS) the rate of mismatches is also much higher, the decrease in the magnitude difference tolerance is an effort to reduce the rate somewhat.

\section{Proper Motion Calculation}\label{method}
	The proper motion calculation method remains almost identical to the method used for the UKIDSS Large Area Survey (LAS) described by \citet{smith14}, with a few alterations which were necessary to deal with the higher source density and passband difference.
	
	We select preliminary reference sources that meet the following criteria based on the data in the FITS file catalogues:
	\begin{description}
		\item Stellar image profile at both epochs;
		\item K at both epochs between 12.25 and 17.00;
		\item Ellipticity at both epochs $<$ 0.3; and
		\item K uncertainties at both epochs $<$ 0.1.
	\end{description}
	The minimum and maximum reference star magnitude selection was changed from J$=$16.0 and J$=$19.6 for the LAS to K$=$12.25 and K$=$17.00 for the GPS. The decrease in the upper magnitude limit reflects a shallower 5$\sigma$ detection limit in the GPS K band than the LAS J band, and limits the number of mismatches selected as reference sources since they become more common at fainter magnitudes. The new lower magnitude limit is a conservative cut of slightly saturated sources and also allows us to retain distant luminous stars as reference points. In the LAS and elsewhere at high Galactic latitude the brightest sources are nearby stars with large motions, owing to the small scale height of the Galactic disc. These make poor astrometric reference stars. In the Galactic plane however, the brightest sources are usually very luminous distant stars with relatively small motions which make good reference stars. Hence it was advantageous to reject bright sources as reference stars in the LAS but the opposite is true for the GPS.
	Using these preliminary reference sources initial motions were calculated as described in \citet{smith14} and we then rejected all reference stars with initial motion $>$1$\sigma$. We rejected all frames (WFCAM arrays) that contained fewer than 100 remaining reference sources since inspection indicated that these contained bad data.
	
	We perform a transformation of second epoch array positions to first epoch array positions using one of two second order polynomial transformations. A transformation matrix is calculated either `globally', i.e. using reference sources across each whole frame; or `locally', i.e. using reference sources from only a portion of the frame local to the target. Where possible we use the local transformation (see \citealp{smith14} Figure 2 for a justification of this preference), which selects reference sources from within a radius governed by the local density and distribution of reference sources. More specifically the radius of reference source selection is the smallest radius in which we find at least 5 reference sources in each quadrant around the target, rounded up to the nearest 20". We set a minimum of 300" on this value. If there are insufficient nearby reference sources then we use the global transformation method instead.
	
	These motions, calculated as the residual to the transformation are relative to the reference stars used. Since these stars are also moving (albeit very slowly) the motion calculated is not absolute. For the LAS we were able to account for this by subtracting the calculated median motion of extragalactic sources relative to the reference stars. The \citet{smith14} relative to absolute proper motion correction relied on reasonably trustworthy morphological classification. \citet{lucas08} found that in the GPS the high source density and resultant high frequency of stellar blends mean erroneous galaxy classifications based on a merge of the J, H, and K band classifications are common. In the FITS file data we use here we have morphological classifications based on a single band detection, which is even less reliable. Furthermore, the factor of $\sim$10 increase in the number of sources meant that it was necessary to split the input catalogue into some 20 subsets and run those through the pipeline individually. The \citeauthor{smith14} method of relative to absolute correction involved selecting extragalactic sources from within 3$^{\circ}$ of each target frame, meaning that fields near the edges of the dataset would be corrected using fewer extragalactic sources than fields $>3^{\circ}$ from the edge. Splitting the input catalogue created many more field edges than would be the case when processing the catalogue as a whole. For these reasons we chose not to perform the relative to absolute conversion.
	
	Despite the lack of relative to absolute correction, the relatively large distance to most stars in the GPS and their consequently very small proper motions should provide a zero point that is fairly near to absolute. We calculated sample relative to absolute corrections using the Besan\c{c}on models \citep{robin03} and find that the correction is always $\leq$ 5 $mas~yr^{-1}$ in the area covered by this paper.

\section{Results}\label{results}
	\subsection{Reliability}\label{reliability}
		Due to the high incidence of false high proper motion detections we found visual verification to be essential. To reduce the number of candidates to a manageable quantity for visual verification we adopted the following quality criteria based on tests on a subset of the data spread throughout the plane.
		\begin{description}
			\item $l>$60$^{\circ}$;
			\item Proper motion $>$200 $mas~yr^{-1}$;
			\item Classified as stellar at both epochs;
			\item Ellipticity at both epochs $<$0.3;
			\item K $<~17$ at either epoch; and
			\item Fewer than 10 other candidates in the same 13.65' $\times$ 13.65' frame.
		\end{description}
		In a preliminary test of this selection we found no genuine high proper motion sources in any frame containing greater than 10 candidates meeting those first five criteria. This is due to inclusion of a small number of poor quality frames due to the use of FITS file data which has not been quality controlled. These poor quality frames generate large numbers of spurious detections, so their identification and removal was desirable.
		The above selection produced 5,655 good candidates for genuine high proper motion sources. Approximately 61\% of sources at K $<$ 17 pass the class and ellipticity cuts, and we therefore adopt this figure as an estimate of the completeness of this selection. We note that the region at $l<$60$^{\circ}$ could be investigated if the search is limited to bright but unsaturated stars in the 12$<$K$<$14 magnitude range.
		
\begin{figure}
			\begin{center}
				\begin{tabular}{c}
					\epsfig{file=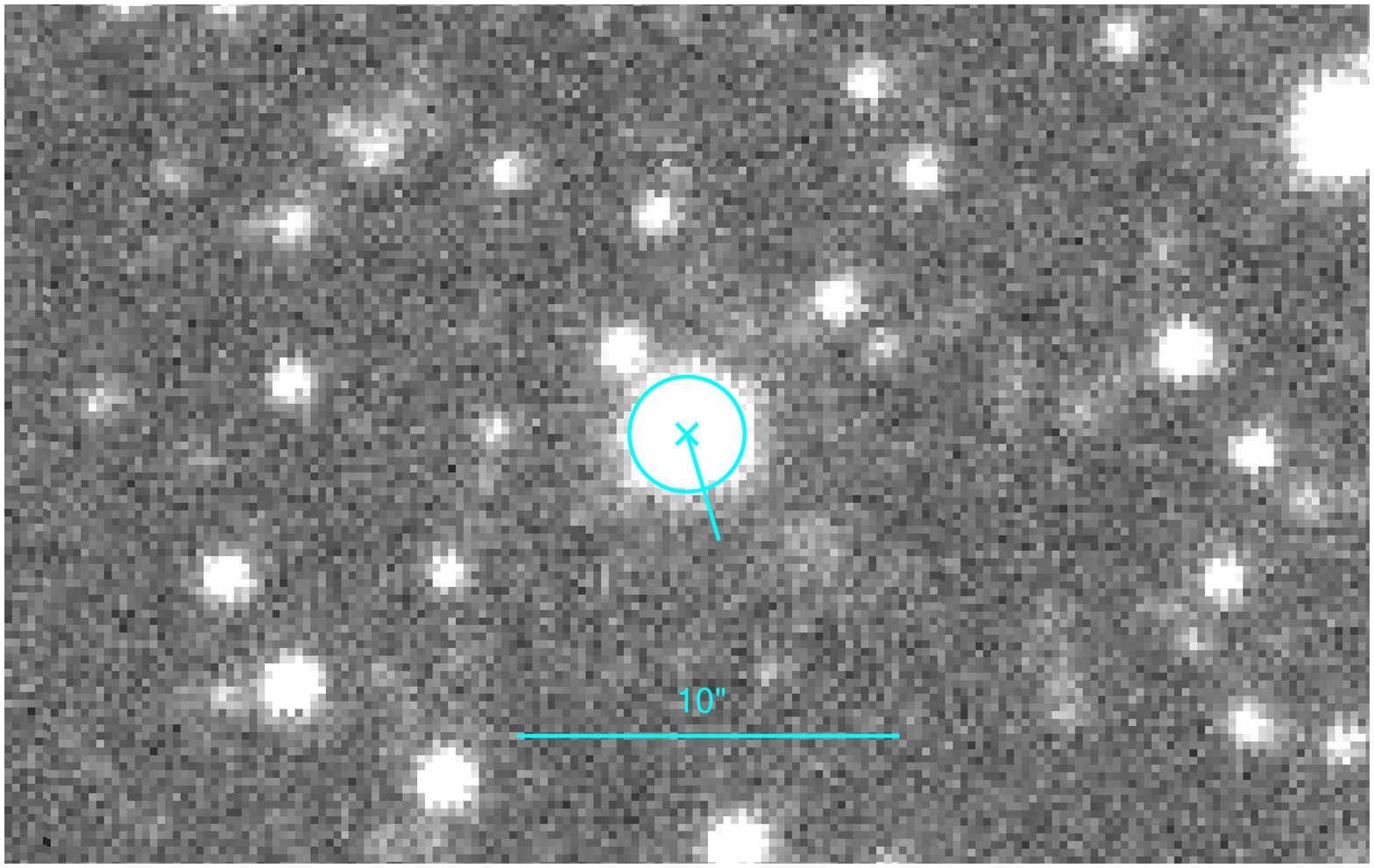,width=1\linewidth,clip=}\\
					\epsfig{file=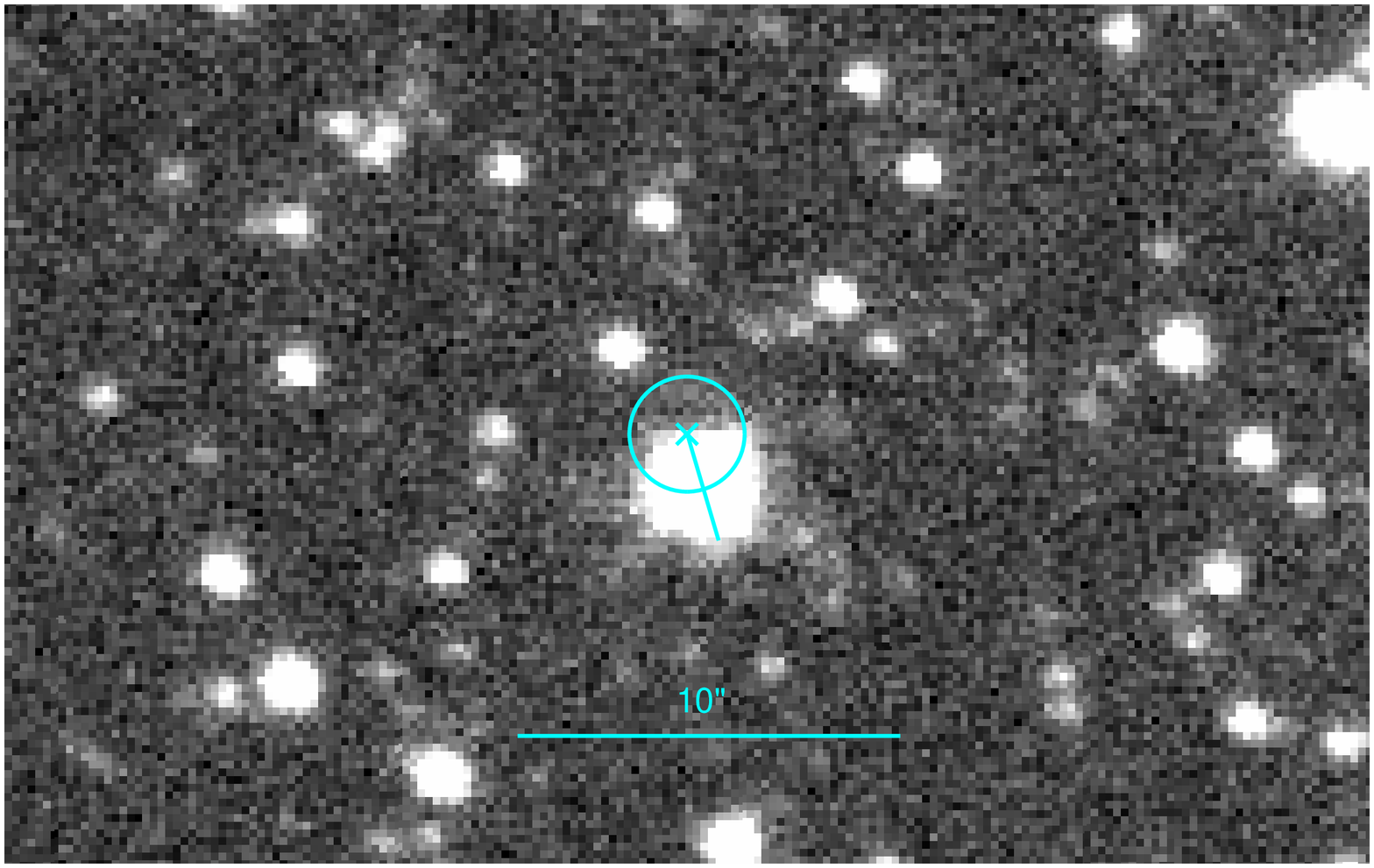,width=1\linewidth,clip=}
				\end{tabular}
				\caption{First epoch K (top) and second epoch K blinking images generated by \textsc{DS9} for the known high proper motion object 2MASS~J19483064+2321473. Our proper motion for this source is 295 $mas~yr^{-1}$. The images are separated by 5.15 years. The regions are placed at the same position on both images relative to their WCS. The $\times$ shows the first epoch K band position of the source, the circle shows the radius of motion between the image epochs, the line shows the direction of travel and its length corresponds to 10 years of motion. The intersection of the circle and line therefore shows the expected position of the source at the second epoch. There is another circle with a radius of 256 pixels centred on the first epoch position of the target, which falls beyond the boundaries of these images, corresponding to the distance of a saturated or near-saturated star that could cause first order cross-talk at the target position. The bar below the source in each image shows 10" and they are oriented north up east left.}
				\label{blinkims}
			\end{center}
		\end{figure}

\begin{figure}
			\begin{center}
				\begin{tabular}{c}
					\epsfig{file=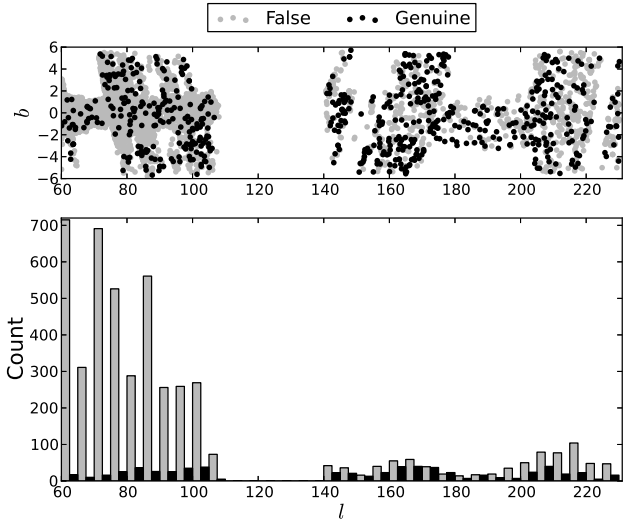,width=1\linewidth,clip=}
				\end{tabular}
				\caption{The distribution in Galactic coordinates of high proper motion candidates identified as false (grey) and genuine (black). The upper panel shows the coverage in Galactic coordinates and that the general distribution of genuine high proper motion objects is fairly uniform across the field, in contrast to the increase in false high proper motion objects with decreasing Galactic longitude. The lower panel shows the distribution in Galactic longitude alone. Increased coverage width in Galactic latitude corresponds to and accounts for the peaks in the distribution, which is otherwise fairly flat. The large number of false high proper motion candidates at low Galactic longitudes is driven by the large increase in source density relative to high Galactic longitudes which increases the frequency of mismatches.}
				\label{genfal_dist}
			\end{center}
		\end{figure}

		To identify genuine high proper motion sources we blinked the candidates in sequence by calling \textsc{DS9} in blink mode. Regions were overlaid showing the position of each source at the first epoch, the calculated position of the source at the second epoch (Fig. \ref{blinkims}), and the radius of first order cross-talk. During array readout electronic cross-talk can cause fainter duplicate images of saturated or near-saturated sources at a distance of 256$n$ pixels (in the case of 2$\times$2 microstepped images as we have here) from the bright source in either the $x$ or $y$ direction depending on the readout direction. \citet{dye06} discuss cross-talk and other data artefacts present in WFCAM data. The overlaid regions made identification of mismatches and cross-talk, which were the dominant source of false detections, straightforward. To add to the visual cross-talk identification we also used software designed to identify possible cross-talk in WFCAM data using the positions of bright nearby 2MASS sources. This identified a small number of additional cross-talk sources that had initially been missed. We identified 617 genuine high proper motion sources from within this sample which gives an overall ratio of false to genuine candidates of 9:1. Figure \ref{genfal_dist} shows the distribution of false and genuine high proper motion detections in Galactic coordinates. The fraction of false positives increases rapidly with decreasing Galactic longitude, the ratio of false to genuine candidates is $\sim$ 1:1 at $l~=~$180$^{\circ}$. The density of genuine high proper motion sources appears fairly uniform.
		
\begin{figure}
			\begin{center}
				\begin{tabular}{c}
					\epsfig{file=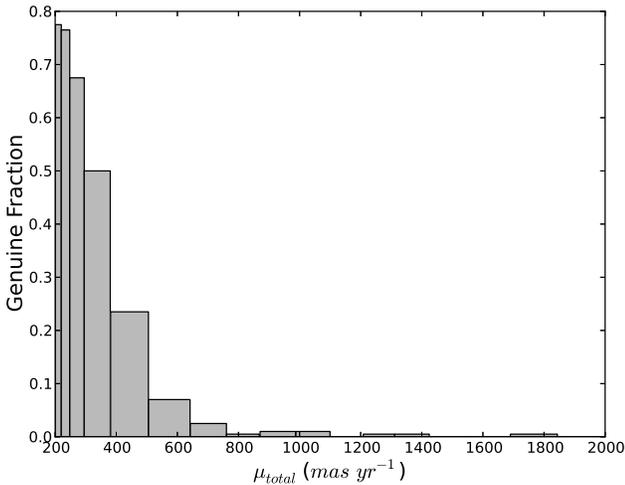,width=1\linewidth,clip=}
				\end{tabular}
				\caption{Histogram showing the decrease in the fraction of genuine high proper motion detections as total proper motion increases. The sample of sources described in Section \ref{reliability} were sorted by total proper motion and binned into groups of 200 sources. Note that the $x$ axis continues to 16 $arcsec~yr^{-1}$, though the genuine fraction remains at zero past what is shown here. The fraction of genuine high proper motion sources increases rapidly with decreasing proper motion.}
				\label{genfal_pm_dist}
			\end{center}
		\end{figure}

		Figure \ref{genfal_pm_dist} shows that the fraction of false positive proper motions is much greater at the high proper motion end. This is unsurprising since the large radius of apparent motion between the images allows for a higher incidence of mismatches. For the same reason sources with an apparent low proper motion are expected to be more reliable than those with high proper motion. At low proper motion the major source of false detections shifts to crosstalk. This can be identified and removed fairly reliably by searching for bright 2MASS sources at a $\sim$51" radius from the target. We find that removal of crosstalk sources in this way increases the genuine fraction amongst relatively low PM candidates (200-300 $mas~yr^{-1}$) to $\sim$85\%.
		
	\subsection{Accuracy}\label{accuracy}
\begin{figure}
			\begin{center}
				\begin{tabular}{c}
					\epsfig{file=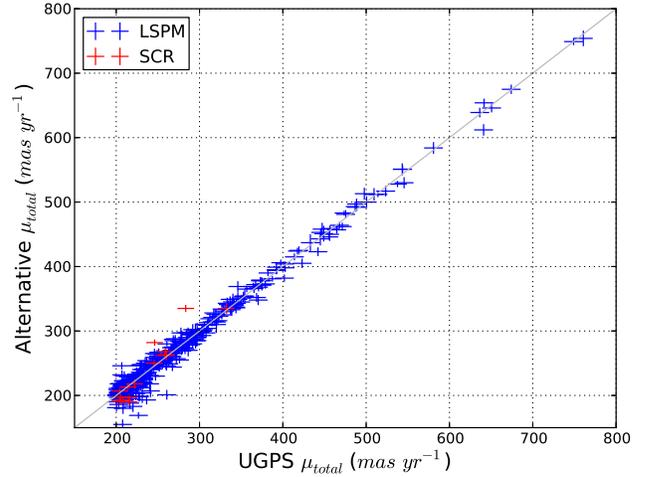,width=1\linewidth,clip=}
				\end{tabular}
				\caption{Our total proper motion compared to those of alternative sources. The alternative sources are given by the legend. The length of the arms of the crosses indicate the uncertainty in the proper motion measurement. LSPM proper motion uncertainties are taken as 8~$mas~yr^{-1}$ in RA and Dec. We estimate the \citeauthor{boyd11b} total proper motion uncertainties at 10~$mas~yr^{-1}$.}
				\label{pm_comparison}
			\end{center}
		\end{figure}

		To evaluate the accuracy of the proper motions we compared them to the long epoch baseline optical catalogues of \citeauthor{lepine05} (\citeyear{lepine05}, LSPM; covering the north) and \citeauthor{boyd11b} (2011a, 2011b; covering the south). We identified 406 sources common to the LSPM catalogue and 15 common to the \citeauthor{boyd11b} catalogues. Figure \ref{pm_comparison} shows a comparison of the total proper motions of the 421 sources between the catalogues, for which Pearson's r correlation coefficient is 0.996. Of the sources in common with the LSPM, 69\% of the proper motions agree within their 1$\sigma$ uncertainties, we omitted the \citeauthor{boyd11b} proper motions from this calculation since the authors do not provide an estimate of their uncertainty. 
		
\begin{figure}
			\begin{center}
				\begin{tabular}{c}
					\epsfig{file=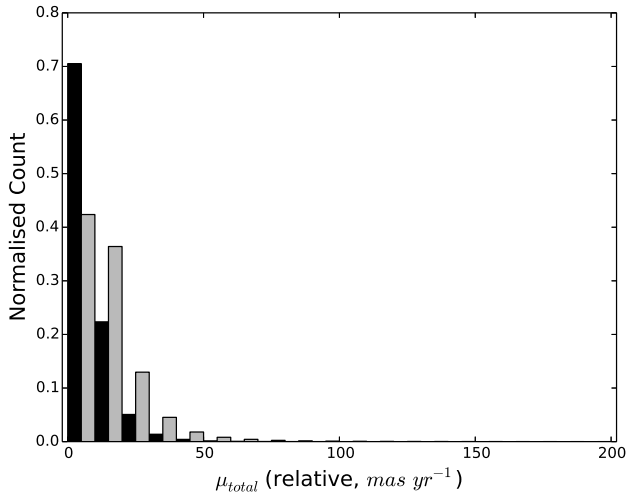,width=1\linewidth,clip=}
				\end{tabular}
				\caption{The distribution of a sample of low proper motion sources in the GPS (black) and the LAS (grey).}
				\label{lowpm_hist}
			\end{center}
		\end{figure}

		To evaluate the accuracy at the lower end of the proper motion scale we selected sources with proper motion $<$~200~$mas~yr^{-1}$, $l$~$>$~60$^{\circ}$, K magnitude at either epoch~$<$~17, K magnitude uncertainty at both epochs~$<$~0.1, ellipticity at both epochs~$<$~0.3, and classified as stellar at both epochs. The distribution of total proper motions of this sample is shown in Figure \ref{lowpm_hist}. For comparison we also include a sample of low proper motion sources from the LAS \citep{smith14}, with J magnitude at either epoch $<$ 18.5. The GPS proper motion distribution is highly peaked towards $\mu~<~$10$~mas~yr^{-1}$ which is expected since we are largely sampling distant main sequence stars and giants with very small proper motion. This tends to suggest that the reliability of the catalogue remains high at $\mu~<~$200$~mas~yr^{-1}$, continuing the trend indicated in Figure \ref{genfal_pm_dist}. 
		By contrast, in the LAS (i.e. outside the Galactic plane) we are sampling relatively nearby stars which might be expected to have (marginally) measurable proper motions, which could explain why the distribution in Figure \ref{lowpm_hist} is less strongly peaked towards zero. We tested this by generating a catalogue sample for the same region and magnitude range as the LAS dataset using the Besan\c{c}on models. We found that the typical proper motions of the model sample were in the range of 0 to 20 $mas~yr^{-1}$ and consistent with the data in Figure \ref{lowpm_hist} after making allowance for the uncertainties in the LAS proper motions.
		
		Note that we still produce proper motions for mildly saturated objects as the morphological classification flag for saturation in the FITS catalogues is not always reliable. As a result we would encourage that proper motions for sources brighter than K$\sim$12 be used with caution.

\section{Discoveries}\label{discoveries}
	In order to characterise the new high proper motion objects we collected r and i band photometry from the INT/WFC Photometric H$\alpha$ Survey (IPHAS, \citealt{drew05}), specifically a preliminary version of the DR2 catalogue \citep{barentsen14}, and the J and H band photometry from either the UKIDSS DR8 GPS catalogue or their FITS file catalogues which were retrieved from the WSA using their archive listing service. Additionally, an IPHAS detection at a third epoch allows us to safely rule out the possibility that two transient objects (e.g. solar system objects) produced an apparent proper motion of a single object.
	
	The r and i band merged catalogues are available in the IPHAS DR2 catalogue. A small number of the high proper motion sources were in IPHAS fields observed in poor weather and for this reason are not present in the DR2 catalogue. However, the photometric uncertainties take the poor observing conditions into account and we included them where necessary. The coordinate of each IPHAS source is taken from the r band. Where an r band detection is missing the i band is used. We took the proper motion of each source into account by identifying the nearest IPHAS source to the GPS source's expected position at the IPHAS epoch. We visually inspected all IPHAS images to ensure that the catalogue photometry was not compromised by blending or astrometric errors. We found 17 matches located $>$ 1" from their expected position, only one of which was a genuine IPHAS match, UGPS~J211859.26+433801.3 (see Sections \ref{newls} and \ref{cataloguecpmsearch}). The other 16 were mismatches usually due to a non-detection of the target in the i band.
	
	We matched to the UKIDSS DR8 GPS catalogue using a radius which took into account the possibility that our position epoch (the first K band epoch), is not always the same as the epoch from which the WSA take positions. We found that 444 of the high proper motion sources were matched to only one UKIDSS DR8 source, with a further 43 instances where there were multiple matches to the same high proper motion source. We believe this to be due to missed matches in the creation of the band-merged catalogues by the WSA since these high proper motion sources almost all have motions between the epochs greater than the 1" matching radius used by the WSA for GPS data. Additionally, the groups are mostly pairs where the first match is a K1 detection only and the other is a K2 detection only (K1 being defined in the WSA as the K epoch contemporaneous with the J and H data).
	
	We then matched to the J and H band FITS file catalogues taking into account the proper motion of the source to calculate the expected position of the target in these images.
	
	Where the catalogue photometry might possibly be unsatisfactory (e.g. an IPHAS non-detection, or a high contrast binary with a small separation where significant PSF overlap was likely to have occured) and the target was deemed interesting (e.g. a UCD candidate), it was necessary to perform additional photometry. We used the \textsc{IRAF} \textsc{DAOPHOT} package in these cases and the targets in question are identified in their tables and/or text.

	\subsection{New High Proper Motion Sources}
		To identify those high proper motion sources already in the literature we cross checked against both SIMBAD and VizieR. Between them these services contain several catalogues of verified high proper motion sources (e.g. the LSPM catalogue, \citealt{lepine05}; the search by \citeauthor{boyd11b}, 2011a, 2011b) which are likely to have previously identified many of the same sources from the 617 that we identified.
		
		We used the SIMBAD script service to compile a list of all stars in their database with proper motion $>$100 $mas~yr^{-1}$. To this we matched the epoch 2000 positions of our high proper motion detections using a 15" matching radius, keeping only the closest match. We considered these matches genuine and the source known if the J, H, and K photometry taken from the 2MASS Point Source Catalogue \citep{skrutskie06} did not differ by more than one magnitude. We note that most matches with differing photometry also had large proper motion differences. We identified 426 of our sources in the SIMBAD database, leaving 191 unknown at this stage.
		We identified all catalogues in the VizieR database which contain any source within 15" of the position of each of our remaining high proper motion candidates. We dismissed identifications from the catalogues which were repeatedly identified but do not contain visually verified high proper motion source discoveries. These include proper motion catalogues such as the USNO-B1.0 Catalog, \citep{monet03}; the PPMXL Catalog \citep{roeser10} and photometric catalogues such as the WISE All-Sky Data Release \citep{wright10}; the UKIDSS-DR6 Galactic Plane Survey \citep{lucas08}. Of the 191 sources checked, 29 were identified in other surveys (the majority identified by \citeauthor{boyd11a}, 2011a, 2011b, which are not present in the SIMBAD database) and 162 had no corresponding sources in any VizieR catalogues other than non-verified proper motion catalogues and single epoch photometric catalogues. Subsequently we identified eight more sources in common with the search by \citet{luhman14a} and another with \citet{luhman14b}. The remaining 153 high proper motion sources we consider to be new discoveries and can be found in Table \ref{resultstable}. We note that the known high proper motion sources include twelve very recent WISE-based proper motion discoveries from \citet{kirkpatrick14}, and \citet{luhman14a} and \citet{luhman14b} which had relatively poor astrometric precision. 
		In Table \ref{WISEPMs} we provide GPS proper motions for these objects, which benefit from higher resolution data and longer time baselines.
		
		\begin{figure}
			\begin{center}
				\begin{tabular}{c}
					\epsfig{file=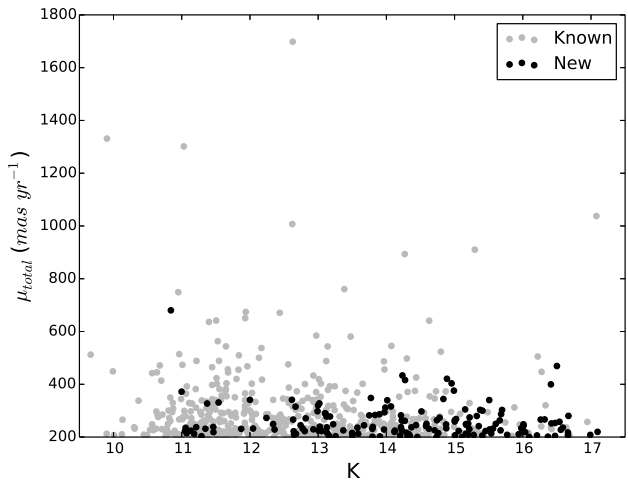,width=1\linewidth,clip=}
				\end{tabular}
				\caption{The distribution of known (grey) and newly discovered (black) high proper motion sources identified by this work.}
				\label{pm_distribution}
			\end{center}
		\end{figure}

	\subsection{New L Dwarf Candidates}\label{newls}
		Brown dwarfs have insufficient mass to support nuclear hydrogen burning. Since they lack a significant internal heating mechanism they cool over time through the brown dwarf sequence. The brown dwarf sequence begins with very young examples of late M type dwarfs and progresses on through L and T type dwarfs into Y dwarfs. An L type dwarf can also be either a low mass star or a brown dwarf depending on its mass and age (see e.g. \citealt{burrows01}). L dwarfs occupy the 2250 K to 1400 K temperature range \citep{kirkpatrick99}, which allows dust to form in their photospheres producing many spectral features not present in warmer objects. Almost a thousand L dwarfs have been identified to date but more discoveries of unusual L dwarfs are still needed to aid the development of evolutionary models and model atmospheres. For example relatively few L dwarfs are known in astrometric binaries, yet these are crucial for accurate mass determination. L dwarfs with unusual colours, low metallicity and halo kinematics, and low surface gravity (see e.g. \citealt{kirkpatrick10}, \citealt{faherty12}) are also poorly sampled.
		
		Since detectable L dwarfs are relatively nearby and as such tend to exhibit large proper motions, we searched our high proper motion sample for previously undetected examples. L dwarfs exhibit a range of near infrared colours dependant on a number of factors. They are relatively faint in optical bandpasses and for this reason optical to infrared colours are very useful for identifying them.
		
		We classified sources as L dwarf candidates if they displayed i-J $>$ 4 (where the IPHAS i magnitude is on the Vega system), or had estimated L spectral types using the system of Skrzypek et al. (in prep). The 13 candidate L dwarf we identified are listed in Table \ref{L_candidates}. One candidate, UGPS J063333.37+100127.4, is unusually blue for an L dwarf, with J$-$K of 0.94. We note however that the IPHAS field containing this object was observed in poor weather, and the photometry is unreliable as a result. This object also falls within the area of the Galactic plane covered by the SDSS. The SDSS i band magnitude of 19.08 gives us an i-J colour of 3.9 which is roughly consistent with a late M dwarf and would match the relatively blue J-K colour. The WISE W1$-$W2 colour for this object is 0.29 $\pm$ 0.05 which is within the range seen for normal late M dwarfs \citep{kirkpatrick11a}. We also note that it is the only candidate in the list with and H$-$K colour greater than its J$-$H colour. Another candidate, UGPS J054457.43+370504.1, is likely to be within the 25 pc volume limited sample. We estimate its distance at 16 to 23 pc assuming a spectral type of L2.5 $\pm$1 subtype and using the spectral type to J band absolute magnitude relation of \citet{dupuy12}. 
		We identified an additional L dwarf candidate which did not meet the original high proper motion candidate selection criteria, UGPS J065420.86+040056.5, see Section \ref{internalcpms}.
		
		\begin{table*}
		\begin{minipage}{110mm}
			\caption{L dwarf candidates based on red optical to infrared colours and a minimum $\chi^2$ fit of the available photometry to brown dwarf colour templates. The i band photometry is from the IPHAS survey and is on the Vega system. We note that these spectral types are approximations based on available photometry, which was limited in some cases.}
			\label{L_candidates}
			\setlength{\linewidth}{.1cm}\newcommand{\contents}{\begin{tabular}{cccccccc}
				\hline
				  \multicolumn{1}{c}{RA} &
				  \multicolumn{1}{c}{Dec} &
				  \multicolumn{1}{c}{K} &
				  \multicolumn{1}{c}{i$-$J} &
				  \multicolumn{1}{c}{J$-$H} &
				  \multicolumn{1}{c}{H$-$K} &
				  \multicolumn{1}{c}{SpTy est.} &
				  \multicolumn{1}{c}{Note} \\
				\hline
				  03:53:04.59 & +47:55:45.6 & 14.59 &         3.8                  & 0.61 & 0.58 &   L0   &  $^a$    \\ 
				  04:46:39.22 & +40:54:52.9 & 15.60 &        $\sim$4.0        & 0.63 & 0.61 &   L1   &  $^d$    \\ 
				  05:44:57.43 & +37:05:04.1 & 12.36 &         4.1                  & 0.80 & 0.74 &  L2.5  &  $^a$    \\ 
				  06:33:33.37 & +10:01:27.4 & 14.21 &         4.1$^f$                  & 0.43 & 0.51 &   L1$^f$   &          \\ 
				  06:47:12.29 & +08:24:40.0 & 15.17 &        $\sim$3.8        & 0.63 & 0.59 &  L0.5  &  $^d$    \\ 
				  19:55:30.48 & +26:13:13.1 & 16.99 &     $>$2.2 $^b$       & 0.82 & 0.67 &  L3.5  &  $^a$    \\ 
				  20:05:30.35 & +36:35:50.7 & 16.01 &        $>$3.0 $^b$    & 0.86 & 0.60 &  L9.5  &  $^d$    \\ 
				  20:17:31.27 & +33:53:59.5 & 16.01 &        $>$3.1 $^b$    & 0.74 & 0.60 &   L2   &  $^d$    \\ 
				  20:31:32.54 & +43:03:19.4 & 14.27 &         4.4                  & 0.83 & 0.75 &   L4   &  $^a$    \\ 
				  21:14:25.17 & +50:10:15.9 & 14.51 &         3.9                  & 0.78 & 0.79 &   L3   &  $^a$    \\ 
				  21:18:59.26 & +43:38:01.3 & 13.50 &   $\sim$3.1 $^c$     & 0.6$^c$ & 0.3$^c$ &        &  $^e$    \\ 
				  21:19:52.83 & +48:26:12.0 & 14.74 &         4.4                  & 0.66 & 0.58 &   L1   &  $^a$    \\ 
				  22:07:06.47 & +53:52:53.3 & 15.00 &         4.1                  & 0.63 & 0.59 &   L1   &  $^a$    \\ 
				\hline
				 \multicolumn{8}{p{\linewidth}}{$^a$ Spectral type based on a photometric typing by Skrzypek et al. (2014, in prep) using available optical and NIR photometry.}\\
				 \multicolumn{8}{p{\linewidth}}{$^b$ Not visible in i band image, lower limit on i band photometry taken as the magnitude of a 3$\sigma$ detection.}\\
				 \multicolumn{8}{p{\linewidth}}{$^c$ Performed own photometry in i, J, H, and K. Near infrared colours have $\pm$0.3 uncertainties.}\\
				 \multicolumn{8}{p{\linewidth}}{$^d$ Spectral type based on a photometric typing by Skrzypek et al. (2014, in prep), though missing several bands.}\\
				 \multicolumn{8}{p{\linewidth}}{$^e$ In a binary or triple system, see Section \ref{newbms}.}\\
				 \multicolumn{8}{p{\linewidth}}{$^f$ Based on IPHAS observations in poor weather, this object has SDSS coverage giving i-J of 3.9 and an revised spectral type estimate of M8.5.}\\
			\end{tabular}}
			\setbox0=\hbox{\contents}
			\setlength{\linewidth}{\wd0}
			\contents
		\end{minipage}
		\end{table*}
	
	\subsection{New Benchmark Ultracool Dwarf Candidates}\label{newbms}
		Ultracool dwarfs as binary companions to other objects (e.g. main sequence stars or white dwarfs) offer an opportunity to test the properties predicted for them by atmospheric models and hence evaluate and refine the models themselves. Age and metallicity are difficult to constrain observationally in UCDs and these properties can sometimes be measured for a companion and then adopted for the UCD since they will usually have formed from the same molecular cloud at a similar time \citep{pinfield06}.
		
		We undertook a search for new benchmark UCD candidates using two methods: The first was a straightforward search of current proper motion catalogues for companions to the 153 previously unidentified high proper motion sources. The second method is a wide search of the full 167 million source results table for common proper motion companions to all 617 genuine high proper motion sources.

		\subsubsection{Existing Catalogue Search}\label{cataloguecpmsearch}
		For the search of existing proper motion catalogues we used the LSPM catalogue for the northern sources, which also contains the Tycho-2 Catalogue of 2.5 Million Bright Stars (Tycho-2, \citealt{hog00}) and the All-sky Compiled Catalogue of 2.5 million stars (ASCC-2.5, \citealt{kharchenko01}), and the catalogue created by \citeauthor{boyd11b} (2011a, 2011b) for the small number of our objects in the south. We performed a 1000" sky match, returning all matches with proper motion difference significance in $\alpha \cos \delta$ and $\delta$ combined $<$ 2$\sigma$. The \citeauthor{boyd11b} search yielded no results. Table \ref{lspmcompanions} shows the seven candidate pairs identified in the LSPM search, two of which are matched to the same GPS source (UGPS~J211859.26+433801.3) and are therefore a candidate triple system. Below we discuss the two systems in which the newly discovered high proper motion object may be an ultracool dwarf.
		
		\begin{table*}
		\begin{minipage}{160mm}
			\caption{The seven candidate companions identified in a search of the LSPM catalogue, two of which are matched to the same GPS source and are therefore together a candidate triple system. The first three columns are those of the GPS source, the next four are of the LSPM candidate companion. The coordinates given are at epoch 2000.0. The proper motions of the candidate companion identified by an asterisk are those of the Tycho-2 catalogue (as indicated in the LSPM catalogue by the astrometric flag), otherwise they are those of the LSPM. All proper motions are in units of $mas~yr^{-1}$. K$_s$ is the 2MASS short K band magnitude obtained from the LSPM catalogue.}
			\label{lspmcompanions}
			\begin{tabular}{|c|c|c|c|l|c|c|r|r|c|}
				\hline
				  \multicolumn{1}{|c|}{$\alpha$} &
				  \multicolumn{1}{c|}{$\delta$} &
				  \multicolumn{1}{c|}{K} &
				  \multicolumn{1}{c|}{i-J} &
				  \multicolumn{1}{c|}{Name} &
				  \multicolumn{1}{c|}{$\mu_{\alpha} \cos \delta$} &
				  \multicolumn{1}{c|}{$\mu_{\delta}$} &
				  \multicolumn{1}{c|}{K$_s$} &
				  \multicolumn{1}{c|}{Separation} &
				  \multicolumn{1}{c|}{$\Delta{}_{\mu}$} \\
				  \multicolumn{8}{c}{} &
				  \multicolumn{1}{c}{(")} &
				  \multicolumn{1}{c}{Significance ($\sigma$)} \\
				\hline
				  04:02:29.42 & +48:12:56.6 & 16.53 & 0.64 & LSPM J0402+4812 & 139 & -223 & 15.49 & 2.1 & 0.66\\ 
				  04:35:19.94 & +43:06:09.4 & 11.13 & 1.94 & LSPM J0435+4305$^{\ast}$ & 155 & -163 & 7.62 & 67.3 & 1.35\\ 
				  21:11:04.39 & +48:00:21.9 & 11.87 & 1.89 & LSPM J2109+4811 & 183 & 127 & 11.25 & 949.2$^a$ & 0.84\\ 
				  21:18:59.26 & +43:38:01.3 & 13.50 & $\sim$3.1 & LSPM J2118+4338 & 177 & 116 & 10.23 & 2.2 & 1.62\\ 
				  21:18:59.26 & +43:38:01.3 & 13.50 & $\sim$3.1 & LSPM J2119+4352 & 170 & 139 & 10.03 & 931.2$^a$ & 0.43\\ 
				  21:32:12.97 & +44:52:29.3 & 15.67 & 1.58 & LSPM J2132+4452 & 164 & 133 & 13.79 & 28.5 & 0.99\\ 
				  21:41:15.07 & +56:40:12.9 & 14.01 & 3.02 & LSPM J2141+5640 & 224 & 247 & 11.34 & 5.6 & 0.62\\ 
				\hline
				  \multicolumn{10}{l}{$^a$ Owing to their very large angular separations these are likely chance alignments, see Section \ref{chancealignnotes}.}\\
			\end{tabular}
		\end{minipage}
		\end{table*}
		
			\paragraph*{}UGPS~J211859.26+433801.3 
				is a close, faint ($\Delta_J$~$\simeq$~3.7, J~$=$~14.7, $\mu$~$=$~215$\pm$4~$mas~yr^{-1}$) companion to the bright M dwarf LP~234-2220. LP~234-2220 was classified as an M3.5 dwarf with an estimated distance of 53.3$\pm$8.1 pc using spectroscopic observations by \citet{reid04}. The pair are separated by 2.2" and the \citet{reid04} spectrum may be contaminated by flux from the companion, though the level of contamination should be small and the spectral classification still reasonably accurate. 
							
				Due to the high contrast and small separation between the pair we deemed it necessary to re-measure the full range of photometry from the original image files. In each case we used a background count level equal to the median counts in a one pixel wide annulus with a radius equal to the separation between the pair centred on LP 234-2220. The uncertainties in the new photometry are relatively large, as much as 0.3 mag in each case. For this reason we relied more on the magnitude of the contrast than optical or infrared colours as an indicator of spectral type since the uncertainty on the contrast is dependant on the uncertainty of only one photometric measurement of the secondary, rather than colour which is dependent on two. A companion i$-$J colour of $\sim$3.1 suggests a spectral type around M6-7. A primary of type M3.5 (M$_J$ $\sim$ 7.8) and a $\Delta_J$ $\simeq$ 3.7 suggests a spectral type for the companion of around M9.5/L0.
			
				We identified another very widely separated (15.5') common proper motion companion to this system: 2MASS~J21193088+4352264 (2MASS~J2119+4352 hereafter, see also Table \ref{internalGPScpmpairs}). The IPHAS r-i colour of 2MASS~J2119+4352 suggests a spectral type of M3.5 based on \citet{drew05} Table 2 and assuming the object is unreddened, which is not an unreasonable assumption given its high proper motion. The similar spectral types and IPHAS optical and GPS infrared magnitudes between LP~234-2220 and 2MASS~J2119+4352 suggest a similar distance. The LSPM proper motions for 2MASS~J2119+4352 and LP~234-2220 which do not suffer from the saturation seen in the GPS for these bright objects, differ by 1.5 $\sigma$. The angular separation gives a projected separation of order 50,000AU at 53pc. Such a system is unlikely to have survived for any significant length of time and the IPHAS narrow band photometry indicates that neither component has any excess H$\alpha$ emission (which would have indicated youth). We find in Section \ref{chancealignnotes} that we expect to find several pairs of sources in our sample with such large angular separations and similar proper motions that are not physically associated. We therefore conclude that the similar proper motions of LP~234-2220 and 2MASS~J2119+4352 are most likely coincidental.
		
			\paragraph*{}UGPS~J214115.07+564012.9 
				(UGPS~J2141+5640B hereafter) is a $\mu$~$=$~339$\pm$12~$mas~yr^{-1}$ common proper motion companion to G 232-30 with a separation of $\sim$5.6". G232-30 appears to be an M0/M1 type dwarf based on its i-J and near infrared colours, which given the J band contrast (2.85) between the pair and assuming their genuine companionship and no unresolved multiplicity in either component suggests that UGPS~J2141+5640B is approximately M5/M6. The i-J colour (3.02) of UGPS~J2141+5640B suggests an approximate spectral type of M6.5 and this is also consistent with its J-H colour (0.57, although this changes very little across the M dwarf sequence). However, H-K~$=$~0.47 for this object, which suggests a later type M dwarf (M7/M8). On balance it is most likely that UGPS~J2141+5640B is an M6.5 dwarf.
		
		\subsubsection{Internal GPS Search}\label{internalcpms}
			For the internal search of the full GPS proper motion results table we search for candidate common proper motion ($\Delta{}_{\mu}~<$ 20 $mas~yr^{-1}$) companions to all 617 identified high proper motion GPS sources with separations up to 30'. We applied no further quality control or brightness selection criteria to the candidate list in order that we not reject any potentially valuable sources due to e.g. a profile misclassification at a single epoch, this selection returned 1032 candidates. A visual inspection yielded 41 genuine high proper motion objects within this sample after removal of 5 duplicate sources from frame overlap regions. Among these we find 11 instances where both components are among the original 617 GPS high proper motion sources and hence they produce a reversed pair (i.e. two instances with switched components), removal of these pairs left us with 19 candidate common proper motion pairs which we show in Table \ref{internalGPScpmpairs}. Since all the companions did not meet the original high proper motion source candidate selection criteria their astrometry is likely compromised and the uncertainty on the proper motion will be underestimated, as a result the stated significance of the proper motion difference should be regarded as a lower limit.
			
			Based on the i-J colours of the original GPS high proper motion sources from Table \ref{internalGPScpmpairs} and their K band contrasts we identified pairs 1, 2, and 12 as candidates for new UCD benchmark objects. Based on the positions of the three candidate primaries in a K band reduced proper motion (H$_K$) vs. i-J plot, the primary in pair 2 appears to be a white dwarf (faint in H$_K$, blue in i-J) while the remaining two candidate primaries appear to be main sequence stars with the primary of pair 12 just on the edge of the subdwarf locus (unremarkable i-J, faint in H$_K$). Inspection of the IPHAS i band images and catalogues showed that the secondary in pair 2 is equal in i band brightness to the primary (17.39) and it is therefore likely that they are a pair of equal mass white dwarfs given their almost identical i-K colour. The secondaries in pairs 1 and 12 are non-detections in the IPHAS i band images and are promising UCD candidates as a result.
			
			\begin{table*}
			\begin{minipage}{176mm}
				\caption{Internal GPS common proper motion companion candidates. Columns 2 to 4 refer to the original GPS high proper motion source and columns 5 to 8 refer to the GPS source which did not meet the original high proper motion candidate selection criteria. The coordinates given are at epoch 2000.0. The dropout note indicates the reason the companion was not selected as an initial high proper motion candidate. Dropout note key: a - flagged as saturated at either epoch; b - flagged as a galaxy at either epoch; c - bad pixel within 2" aperture flag at either epoch; d - ellipticity $>$ 0.3 at either epoch; e - $\mu_{tot}$ just below original selection criteria (200 $mas~yr^{-1}$); f - neither epoch K band magnitudes below 17. In the `Known' column the left tick/cross corresponds to the original GPS source, the right tick/cross corresponds to the new GPS companion candidate.}
				\label{internalGPScpmpairs}
				\begin{tabular}{|c|c|c|r|c|c|c|r|c|c|r|c|c|}
					\hline
					  \multicolumn{1}{c}{Pair} &
					  \multicolumn{1}{|c|}{$\alpha$} &
					  \multicolumn{1}{c|}{$\delta$} &
					  \multicolumn{1}{c|}{K} &
					  \multicolumn{1}{c|}{} &
					  \multicolumn{1}{c|}{$\alpha$} &
					  \multicolumn{1}{c|}{$\delta$} &
					  \multicolumn{1}{c|}{K} &
					  \multicolumn{1}{c|}{Dropout} &
					  \multicolumn{1}{c|}{} &
					  \multicolumn{1}{c|}{Separation} &
					  \multicolumn{1}{c|}{$\Delta{}_{\mu}$} &
					  \multicolumn{1}{l|}{Known} \\
					  \multicolumn{1}{c}{} &
					  \multicolumn{1}{c}{} &
					  \multicolumn{1}{c}{} &
					  \multicolumn{1}{c}{} &
					  \multicolumn{1}{c}{} &
					  \multicolumn{1}{c}{} &
					  \multicolumn{1}{c}{} &
					  \multicolumn{1}{c}{} &
					  \multicolumn{1}{c}{Note} &
					  \multicolumn{1}{c}{} &
					  \multicolumn{1}{c}{(")} &
					  \multicolumn{1}{c}{Significance ($\sigma$)} &
					  \multicolumn{1}{l}{} \\
					\hline
					  1 & 03:42:14.85 & +54:10:19.6 & 11.14 &  & 03:42:14.52 & +54:10:18.8 & 15.69 & bd & & 2.9 & 1.10 &  \ding{51} \ding{55}$^{~}$   \\
					  2 & 04:02:29.42 & +48:12:56.6 & 16.54 &  & 04:02:29.21 & +48:12:56.8 & 16.57 & b & & 2.1 & 0.47 &  \ding{55} \ding{55}$^1$   \\
					  3 & 04:33:14.79 & +40:47:35.8 & 12.63 &  & 04:33:14.76 & +40:47:37.8 & 13.02 & b & & 2.0 & 0.10 &  \ding{51} \ding{55}$^1$   \\
					  4 & 04:39:15.47 & +39:06:31.1 & 13.82 &  & 04:39:16.15 & +39:06:36.4 & 12.52 & e & & 9.4 & 1.48 &  \ding{51} \ding{51}$^{~}$   \\
					  5 & 05:15:07.12 & +45:13:00.6 & 11.80 &  & 05:15:06.99 & +45:12:59.9 & 12.84 & b & & 1.6 & 0.93 &  \ding{51} \ding{55}$^1$   \\
					  6 & 05:21:43.70 & +33:22:05.5 & 10.01 &  & 05:21:43.46 & +33:21:58.8 & 15.70 & b & & 7.4 & 1.15 &  \ding{51} \ding{51}$^{~}$   \\
					  7 & 05:27:46.10 & +44:35:34.2 & 11.58 &  & 05:27:46.99 & +44:36:11.4 & 9.62 & a & & 38.4 & 0.39 &  \ding{51} \ding{51}$^{~}$   \\
					  8 & 05:27:46.94 & +44:36:13.9 & 10.87 &  & 05:27:46.99 & +44:36:11.4 & 9.62 & a & & 2.7 & 1.05 &  \ding{51} \ding{51}$^{~}$   \\
					  9 & 05:34:21.85 & +42:27:39.0 & 11.91 &  & 05:34:21.82 & +42:27:37.3 & 13.29 & b & & 1.8 & 0.68 &  \ding{51} \ding{55}$^1$   \\
					  10 & 05:47:33.08 & +38:03:05.5 & 15.02 &  & 05:47:32.93 & +38:03:05.2 & 16.91 & b & & 1.8 & 0.44 &  \ding{55} \ding{55}$^{~}$   \\
					  11 & 06:48:20.56 & +05:40:33.5 & 11.20 &  & 06:48:20.30 & +05:40:30.1 & 9.09 & a & & 5.1 & 2.12 &  \ding{51} \ding{51}$^{~}$   \\
					  12 & 06:53:11.66 & +03:47:50.7 & 12.44 &  & 06:54:20.86 & +04:00:56.5 & 17.11 & cf & & 1300.1$^2$ & 0.95 &  \ding{51} \ding{55}$^{~}$   \\
					  13 & 20:40:04.51 & +42:21:07.1 & 12.03 &  & 20:39:52.64 & +42:20:33.8 & 10.91 & d & & 135.7 & 0.64 &  \ding{51} \ding{51}$^{~}$   \\
					  14 & 20:55:51.57 & +43:27:48.1 & 13.22 &  & 20:56:56.34 & +43:08:08.4 & 11.19 & a & & 1375.4$^2$ & 0.84 &  \ding{51} \ding{51}$^{~}$   \\
					  15 & 21:18:59.26 & +43:38:01.3 & 13.50 &  & 21:18:59.30 & +43:38:03.5 & 10.04 & a & & 2.2 & 1.83 &  \ding{55} \ding{51}$^{~}$   \\
					  16 & 21:18:59.26 & +43:38:01.3 & 13.50 &  & 21:19:30.94 & +43:52:26.8 & 9.89 & a & & 931.2$^2$ & 1.27 &  \ding{55} \ding{51}$^{~}$   \\
					  17 & 21:23:42.21 & +44:19:17.1 & 12.76 &  & 21:23:43.45 & +44:19:28.0 & 10.59 & a & & 17.2 & 0.13 &  \ding{51} \ding{51}$^{~}$   \\
					  18 & 22:21:29.14 & +55:56:00.1 & 14.20 &  & 22:23:34.94 & +56:10:05.1 & 10.39 & a & & 1350.9$^2$ & 1.61 &  \ding{55} \ding{51}$^{~}$   \\
					  19 & 22:37:06.22 & +55:54:40.8 & 10.82 &  & 22:37:05.95 & +55:54:44.3 & 9.33 & a & & 4.2 & 0.55 &  \ding{51} \ding{51}$^{~}$   \\
					\hline
					\multicolumn{13}{l}{$^1$  Known high proper motion object is actually a blend of both components, previously unresolved.} \\
					\multicolumn{13}{l}{$^2$ Owing to their very large angular separations these are likely chance alignments, see Section \ref{chancealignnotes}.}\\
				\end{tabular}
			\end{minipage}
			\end{table*}

			\paragraph*{}UGPS~J034214.85+541019.6 AB: 
				- pair 1 in Table \ref{internalGPScpmpairs}, UGPS~J0342+5410 AB hereafter. UGPS~J0342+5410 B is an $\mu$~$=$~240$\pm$12~$mas~yr^{-1}$ IPHAS i band non-detection, the 3$\sigma$ detection limit of the field is 21.2. The pair are separated by 2.9". Flux from the primary at this radius only increases the background count level by of order 30\% in the i band image so the detection limit should still be reasonably accurate. UGPS~J0342+5410 B has J$=$16.67$\pm$0.02 and H$=$16.12$\pm$0.02, this gives a lower limit on i-J of 4.5 which corresponds to a spectral type later than or equal to approximately L1. The infrared colours of UGPS~J0342+5410 B (J-H $=$ 0.55, H-K $=$ 0.43) are somewhat bluer than expected for a typical L type dwarf and suggest a late M spectral type. The near infrared colours of UGPS~J0342+5410 A suggest a spectral type of approximately M1.5 ($\pm$ 1.5 subtypes, using 2MASS photometry as the GPS photometry is near saturation). Given the contrast ratio between the pair we can say the secondary is between M8 and L1 and could be either a red late M dwarf or a blue early L dwarf, follow up spectroscopy will be needed for confirmation.
			
			\paragraph*{}UGPS~J065311.66+034750.7 and UGPS~J065420.86+040056.5: 
				- pair 12 in Table \ref{internalGPScpmpairs}, UGPS~J0653+0347 ($\mu$~$=$~251$\pm$6~$mas~yr^{-1}$) and UGPS~J0654+0400 ($\mu$~$=$~258$\pm$15~$mas~yr^{-1}$) hereafter. UGPS~J0654+0400 is an IPHAS i band non-detection, the 3$\sigma$ detection limit of the field is 20.3. It has J and H band magnitudes of 18.86$\pm$0.07 and 17.86$\pm$0.05 respectively giving a lower limit on i-J of approximately 1.4. The near infrared colours of UGPS~J0654+0400 (J-H $=$ 1.00, H-K $=$ 0.75) indicate it is an L dwarf (see e.g. \citealt{dayjones13}). The \citet{dupuy12} spectral type to absolute magnitude relation for L0 to L9 dwarfs suggest a distance to UGPS~J0654+0400 of 75 to 300 pc. This pair are separated by 21.7', and we find in Section \ref{chancealignnotes} that we expect several pairs of sources in our sample with such large angular separations and similar proper motions that are not physically associated and therefore conclude that this likely to be a chance alignment.
				
		\subsubsection{Notes on Chance Alignment Probabilities}\label{chancealignnotes}
			Here we evaluate the probability that the common proper motion companions discussed above are chance alignments. Our estimates are drawn from large simulations made with the online Besan\c{c}on synthetic stellar population tool \citep{robin03}. The catalogues were generated with a K magnitude range equivalent to our our high proper motion sample, using the Galactic position of each candidate. We generated "small field" simulated catalogues with several million stars (equivalent to a 1500~deg$^2$ area but with properties fixed for the precise Galactic location) for each of the binary candidates discussed. The catalogue simulations generate realistic proper motions for each source but do not produce physically associated systems (such as moving groups or binaries). All common proper motion companions in the sample are therefore purely chance alignments. We then identified all sources within each simulated catalogue that have a proper motion consistent with the GPS component of the binary candidate, within the 2$\sigma$ uncertainty on the proper motion difference. The approximate probability of a source appearing within a given angular separation $r$ and with a common proper motion to one of our GPS high proper motion sources is therefore the number of matches in the simulated catalogue multiplied by the area of a circle with radius $r$, divided by 1500 deg$^2$. We must also take into account that we have 617 high proper motion objects, and therefore 617 chances of finding such a chance alignment. For each binary candidate discussed above we treated this using a simple multiplicative factor. In fact there is some variation in the number of proper motion matches in the simulated catalogues, depending on Galactic coordinates and the direction of proper motion but tests indicate that this is less than a factor of 4. In the cases discussed above with separations $<$ 1' the probability is always less than 10$^{-3}$ so the factor of 4 is not significant. Note that for each candidate companion the Galactic coordinates, proper motions and angular separations are different and they require a unique calculation as a result.
			
In Tables \ref{lspmcompanions} and \ref{internalGPScpmpairs} we listed six very widely separated ($>$10') common proper motion companions to sources in our sample, two of which contain a UCD candidate that were discussed in more detail. For these two systems the above calculation indicates that the expected number of chance alignments in our sample is 8 and 13 respectively. Even allowing for the factor of 4 uncertainty we still expect to find a few examples in our data so we conclude that these are chance alignments.

	\subsection{New T Dwarfs}\label{newTs}
		T type brown dwarfs are later in the brown dwarf sequence than L dwarfs. As they age and cool through the brown dwarf sequence the dust present in L dwarf atmospheres sinks below the photosphere and its effect on their spectra disappears. The cooler temperatures of T dwarfs allow molecules such as methane and water to form, which are responsible for the deep absorption features observed in their spectra. They emit most of their radiation in the near infrared and for this reason recent large scale near infrared sky surveys (e.g. 2MASS, \citealt{skrutskie06}; DENIS, \citealt{epchtein97}; CFBDS, \citealt{delorme10}; UKIDSS, \citealt{lawrence07}) are responsible for the majority of current T dwarf discoveries. Later type T dwarfs are cooler still and the WISE mission \citep{wright10} in the mid infrared becomes more sensitive to them at around T6 and later. To date several hundred T dwarfs have been identified.
		
		T dwarfs are extremely faint, even in the near infrared and are only detectable by the current generation of large scale NIR surveys out to of order 100pc. Due to the close proximity of detectable T dwarfs they tend to exhibit relatively large proper motions. For this reason proper motion searches such as this could be expected to identify many examples of T dwarfs. We have found two new examples of T dwarfs, which we describe below, among the 153 previously unidentified high proper motion sources due to their characteristic blue J-H and H-K colors. We also recover UGPS J0722-05 amongst the 617 high proper motion sources. The two GPS T dwarfs identified by \citet{burningham11} are fainter than our K $=$ 17 cut and were not recovered as a result. Two very nearby bright T dwarfs in the GPS footprint were identified by the WISE team recently. WISE~J192841.35+235604.9 \citep{mace13a} lies outside the area covered by this paper and WISE~J200050.19+362950.1 \citep{cushing14} was excluded due to a high ellipticity and a profile misclassification in the second epoch K band observation. Our proper motion for WISE~J200050.19+362950.1 is 75$\pm$7 and 415$\pm$9 $mas~yr^{-1}$ in $\alpha \cos \delta$ and $\delta$ respectively.
		
		\paragraph*{} 
			UGPS~J20480024+503821.9 (UGPS~J2048+5038 hereafter) was identified as a J$=$16.30 $\mu$~$=$~267$\pm$11~$mas~yr^{-1}$, IPHAS i band non-detection. The 3$\sigma$ IPHAS i band detection limit of this field is 20.4$\pm$0.4, which is consistent with approximately M9 or later based on i-J ($>$4.1). The blue nature of this object in J$-$H and H$-$K (-0.07 and 0.05 respectively) and the H$-$W2 and W1$-$W2 colours ($\sim$1.5 and $\sim$1.6 respectively) suggested that this was likely a mid T type dwarf. The source is not included in the WISE All Sky Catalogue nor the AllWISE catalogue but it is visible in the WISE images as a faint source in a crowded field and it is included in the WISE L1b source table, which includes less reliable photometry. The H$-$W2 and W1$-$W2 colours ($\sim$1.5 and $\sim$1.6 respectively) are bluer than expected for a mid T dwarf but the uncertainties should be assumed to be large.
			
\begin{figure}
				\begin{center}
					\begin{tabular}{c}
						\epsfig{file=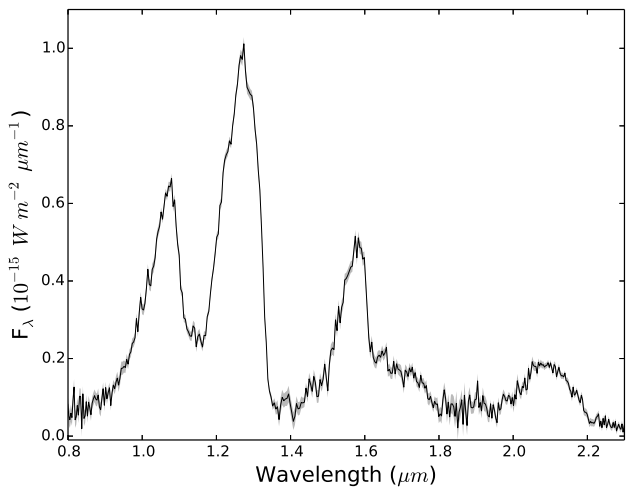,width=1\linewidth,clip=}
					\end{tabular}
					\caption{An IRTF SpeX spectrum of the previously unidentified T5 dwarf UGPS~J2048+5038.}
					\label{UGPS2048_spec}
				\end{center}
			\end{figure}

\begin{figure}
				\begin{center}
					\begin{tabular}{c}
						\epsfig{file=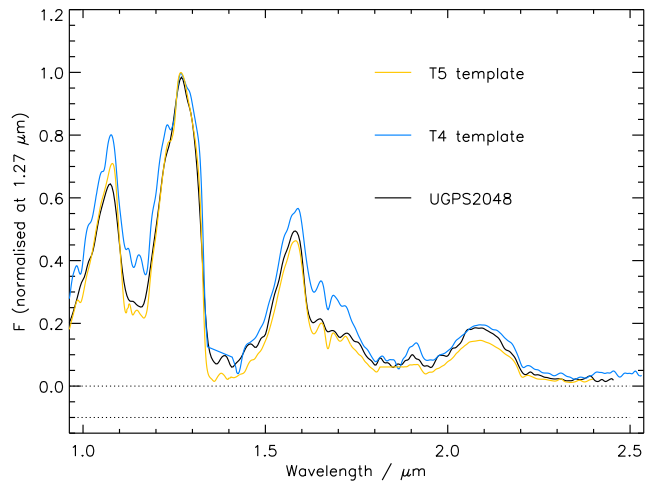,angle=90,width=1\linewidth,clip=}
					\end{tabular}
					\caption{Our $YJHK$ spectrum of UGPS~J2048+5038 compared to the T4 (2MASS~J15031961+2525196 and T5 (2MASS~J15031961+2525196) spectral templates defined in \citet{burgasser06}.}
					\label{fig:sptype}
				\end{center}
			\end{figure}

			We obtained a NASA Infrared Telescope Facility (IRTF) SpeX \citep{rayner03} spectrum of UGPS~J2048+5038 (see Figure \ref{UGPS2048_spec}) on the 30th of September 2013 using the 0.8" slit in prism mode with six AB nod cycles of 200s per nod, although the target drifted out of the slit for the final two and these were discarded as a result. This gave a total on source time of 1600s. We also observed the spectral standard HD199217 for use in removing telluric features from the target spectrum and flux calibration. We combined and reduced the spectrum using the standard reduction tool: SpeXTool \citep{cushing04}. \\
			We have classified UGPS~J2048+5038 following the spectral typing scheme laid out by \citet{burgasser06} for T dwarfs. In Figure \ref{fig:sptype} we show our SpeX YJHK spectrum of UGPS~J2048+5038 compared to the T4 (2MASS~J22541892+3123498) and T5 (2MASS~J15031961+2525196) spectral templates of \citet{burgasser06}. The new source matches the T5 template very closely in the $J$ and $H$ band flux peaks, but shows less flux in the $Y$ band peak, and enhanced flux in the $K$ band.  In Table~\ref{tab:spratios} we give the spectral flux ratios used for index based classification in the \citet{burgasser06} scheme. These values further support the classification of UGPS~J2048+5038 as a T5, and we thus adopt this classification for this object (T5$\pm$0.5). See Table \ref{Tdwf_info} for a list of parameters. The \citet{dupuy12} spectral type to MKO J band absolute magnitude relation gives an $M_J$ of 14.44 for an isolated T5 dwarf, which puts UGPS~J2048+5038 (J$=$16.30) at a distance of approximately 24pc.
				
			\begin{table}\addtolength{\tabcolsep}{-1pt}
				\caption{The spectral flux ratios used for classifying UGPS~J2048+5038.}
				\label{tab:spratios}
				\begin{center}
				\begin{tabular}{c c c c }
					  \hline
					 Index & Ratio & Value & Type \\
					\hline
					H$_2$O-J & $\frac{\int^{1.165}_{1.14} f(\lambda)d\lambda}{\int^{1.285}_{1.26}f(\lambda)d\lambda }$ & $0.26 \pm 0.01$  & T5 \\[10pt]
					CH$_4$-J & $\frac{\int^{1.34}_{1.315} f(\lambda)d\lambda}{\int^{1.285}_{1.26}f(\lambda)d\lambda }$ &  $0.41 \pm 0.01$ & T5 \\[10pt]
					H$_2$O-$H$ & $\frac{\int^{1.52}_{1.48} f(\lambda)d\lambda}{\int^{1.60}_{1.56}f(\lambda)d\lambda }$ & $0.38 \pm 0.01$  & T4/5 \\[10pt]
					CH$_4$-$H$ & $\frac{\int^{1.675}_{1.635} f(\lambda)d\lambda}{\int^{1.60}_{1.56}f(\lambda)d\lambda }$ & $0.42 \pm 0.01$ & T5 \\[10pt]
					CH$_4$-K &  $\frac{\int^{2.255}_{2.215} f(\lambda)d\lambda}{\int^{2.12}_{2.08}f(\lambda)d\lambda }$ & $0.21 \pm 0.01$ & T5 \\[10pt]
					\hline
				\end{tabular}
				\end{center}
			\end{table}
			
			\paragraph*{} 
				An initial IPHAS DR2 match to UGPS~J03553200+4743588 (UGPS~J0355+4743 hereafter, $\mu$~$=$~469$\pm$16~$mas~yr^{-1}$) found the closest match to have a separation $>$1" of the expected position of the target at the IPHAS epoch. Subsequent visual inspection of the IPHAS i band image confirmed a mismatch or blend with a background source. We attempted a subtraction of the PSF of the background source using the standard \textsc{IRAF} \textsc{allstar} program and subsequent inspection of the residual image showed no remaining sign of either the background source or the target. This suggested either the pair were so close they were completely unresolved, or more likely given the expected $\sim$1.1" (3.3 pixel) separation at the IPHAS epoch, the target is simply too faint in the i band to be detected.
			
				An approximate 3$\sigma$ IPHAS i band detection limit of this field is 20.5 magnitudes, suggesting the target is approximately L0 or later type. Given the blue nature of UGPS~J0355+4743 in the near infrared, (J-H~$=$~-0.37 and H-K~$=$~0.08), and H$-$W2 and W1$-$W2 of $\sim$2.67 and $\sim$1.12 respectively we regard it as a bona fide T dwarf and with an estimate type of T6 type dwarf.
			
				The \citet{dupuy12} spectral type to MKO J band absolute magnitude relation gives an $M_J$ of 14.78 for an isolated T6 dwarf, which puts UGPS~J0355+4743 (J $=$ 16.20) at a distance of approximately 19pc. If the target is an unresolved binary or system of higher order multiplicity then this distance is an underestimate.
			
				\begin{table}
					\caption{Parameters of the two previously unidentified T dwarfs which we describe in Section \ref{newTs}.}
					\label{Tdwf_info}
					\begin{center}
					\begin{tabular}{|l|l|l|}
						\hline
						   & UGPS~J2048+5038 & UGPS~J0355+4743 \\
						\hline
						  Right Ascension & 20:48:00.24 & 03:55:32.00\\
						  Declination & +50:38:21.9 & +47:43:58.8\\
						  Spectral Type & T5 & $\sim$T6\\
						  J & 16.30 & 16.20\\
						  $\mu{}_{total}$ & 267 $mas~yr^{-1}$ & 469 $mas~yr^{-1}$\\
						  Distance & $\sim$24 $pc$ & $\sim$19 $pc$\\
						  J$-$H & -0.07 & -0.37\\
						  H$-$K & 0.05 & 0.08\\
						  W1$-$W2 & $\sim$1.6 & 1.12\\
						  H$-$W2 & $\sim$1.5 & 2.67\\
						\hline
					\end{tabular}
					\end{center}
				\end{table}
				
	\subsection{Other Objects of Note}
		\paragraph*{}UGPS~J04514383+4549580 ($\mu$~$=$~680$\pm$7~$mas~yr^{-1}$) is a faint companion to LHS~1708, a G1 type main sequence star, with a separation of 5.6" and a $\Delta_J$ of 5.9 magnitudes. It was identified in our search for new candidate UCD benchmark objects. Our proper motion differs by 2.1$\sigma$ from the Hipparcos proper motion of LHS~1708 and as a result the pair did not show in the $<$~2$\sigma$ candidate companion list in Section \ref{cataloguecpmsearch}. Given the K band brightness (10.8), the uncertainty on our proper motion for UGPS~J04514383+4549580 ($\pm$~7~$mas~yr^{-1}$) is likely underestimated by the pipeline and the significance of the similarity in the proper motion of this pair is therefore also underestimated. Based on its IPHAS optical and GPS infrared colours UGPS~J04514383+4549580 is either a mid-M type main sequence star or a white dwarf.
		LHS~1708 does have an entry in the Washington Double star Catalogue (WDS, \citealt{mason01}), although the stated separations (101.6" and 85.80" for the 1909 and 1989 epochs respectively) are higher than the object we have identified. The position of the WDS secondary at the 1909 and 1989 epochs, which we calculated from the stated separations and position angles relative to the position of LHS~1708 at the two epochs, is consistent with a bright source on the GPS image which shows no proper motion. We conclude that the secondary given in the WDS does not share a proper motion with LHS~1708 and is therefore not a genuine companion. However, our object UGPS~J04514383+4549580, is a genuine common proper motion companion to LHS~1708.

\section{Summary}\label{summary}
	We present the results of a search for high proper motion objects in the UKIDSS Galactic Plane Survey. We selected 5,655 high proper motion ($\mu$~$>$~200~$mas~yr^{-1}$) candidates from 900 deg$^2$ of sky at $l~>$~60$^{\circ}$ and K~$<$~17 for visual verification and found 617 to be genuine, 153 of which were previously unidentified. Among the new high proper motion discoveries we identified two new mid T dwarfs that are likely to be within 25 pc, a further thirteen new L dwarf candidates and two ultracool dwarf binary candidates.
	
	The large 24" matching radius we adopted in an effort to detect objects with very high proper motions at the expense of a large number of mismatches gave an overall ratio of false to genuine candidates of 9:1. At high galactic longitudes, where the source density is much lower, we found this to be less of a problem; the ratio of false to genuine candidates at $l=$180$^{\circ}$ is $\sim$1:1. The accuracy of the genuine high proper motions is good, the median uncertainty on the proper motions of sources at K~$<$~16 is 6.6 $mas~yr^{-1}$. Our proper motions for sources in common with existing long epoch baseline optical catalogues are in good agreement within their uncertainties.
	
	Proper motions were calculated for 167 million sources in total and we plan to extend our search to objects with lower but not insignificant motions. We also plan to extend the selection presented here to data taken after March 31st 2013, search for brighter high proper motion objects at $l$~$<$~60$^{\circ}$, and search for high proper motion objects at K~$>$~17 with the aid of colour selections.

\section*{Acknowledgements}
We would like to thank the referee, John Gizis, for a very positive review.
LS acknowledges a studentship funded by the Science \& Technology Facilities Research Council (STFC) of the UK; PWL, PP, DJP, GB, JED acknowledge the support of a consolidated grant (ST/J001333/1) also funded by STFC.
This work is based in part on data obtained as part of the UKIRT Infrared Deep Sky Survey.
The authors would like to acknowledge the Marie Curie 7th European Community Framework Programme grant n.247593 Interpretation and Parameterization of Extremely Red COOL dwarfs (IPERCOOL) International Research Staff Exchange Scheme.
A.H.A. acknowledges CNPq grant PQ306775/2009-3. 
D.R.R. acknowledges support from FONDECYT grant 3130520. 
This research has made use of the SIMBAD database and VizieR catalogue access tool, operated at CDS, Strasbourg, France.
This research has made use of NASA's Astrophysics Data System Bibliographic Services.
This research has made use of SAOImage \textsc{DS9}, developed by Smithsonian Astrophysical Observatory.
This research has made use of the SpeX Spectrograph and Imager on the NASA Infrared Telescope Facility, see \citet{rayner03}.
The authors would like to acknowledge Setpoint Hertfordshire and the Nuffield Foundation for organising and funding the research placement of R. Bunce.

\bibliographystyle{mn2e}

\appendix\section{}
	\begin{table*}
\begin{minipage}{155mm}
\caption{The high proper motion sources identified for the first time in this publication. The coordinates given are at epoch 2000.0}\label{resultstable}
\begin{tabular}{|l|l|l|l|l|l|l|l|l|}
\hline
  \multicolumn{1}{|c|}{$\alpha$} &
  \multicolumn{1}{c|}{$\delta$} &
  \multicolumn{1}{c|}{$\mu_{\alpha} \cos \delta$} &
  \multicolumn{1}{c|}{$\mu_{\delta}$} &
  \multicolumn{1}{c|}{r} &
  \multicolumn{1}{c|}{i} &
  \multicolumn{1}{c|}{J} &
  \multicolumn{1}{c|}{H} &
  \multicolumn{1}{c|}{K} \\
\hline
  03:20:43.87 & +59:18:23.2 & 146$\pm$7 & -137$\pm$9 &      & 19.64$\pm$0.14 & 16.20$\pm$0.02 & 15.60$\pm$0.02 & 15.11$\pm$0.02 \\
  03:28:47.25 & +59:10:56.0 & 176$\pm$5 & -102$\pm$7 & 18.64$\pm$0.03 & 16.89$\pm$0.02 & 15.08$\pm$0.02 & 14.63$\pm$0.02 & 14.31$\pm$0.02 \\
  03:39:05.89 & +59:35:42.0 & 318$\pm$6 & -294$\pm$6 &      & 19.26$\pm$0.11 & 15.34$\pm$0.02 & 14.79$\pm$0.02 & 14.23$\pm$0.02 \\
  03:46:24.92 & +54:36:26.5 & 128$\pm$6 & -162$\pm$7 & 18.78$\pm$0.02 & 17.52$\pm$0.02 & 15.85$\pm$0.02 & 15.27$\pm$0.02 & 15.01$\pm$0.02 \\
  03:46:57.15 & +56:40:33.5 & 17$\pm$5 & -225$\pm$5 &      & 17.91$\pm$0.05 & 15.04$\pm$0.02 & 14.63$\pm$0.02 & 14.23$\pm$0.02 \\
  03:49:02.69 & +55:34:19.8 & 186$\pm$7 & -235$\pm$8 & 19.49$\pm$0.03 & 17.98$\pm$0.02 & 16.25$\pm$0.02 & 15.71$\pm$0.02 & 15.41$\pm$0.02 \\
  03:53:04.59 & +47:55:45.6 & -81$\pm$6 & -237$\pm$5 &      & 19.54$\pm$0.18 & 15.78$\pm$0.02 & 15.17$\pm$0.02 & 14.59$\pm$0.02 \\
  03:55:31.53 & +47:44:00.6 & 438$\pm$12 & -168$\pm$11 &      &      & 16.20$\pm$0.02 & 16.58$\pm$0.02 & 16.50$\pm$0.04 \\ 
  04:02:29.42 & +48:12:56.6 & 127$\pm$10 & -224$\pm$10 & 18.00$\pm$0.02 & 17.46$\pm$0.02 & 16.82$\pm$0.02 & 16.57$\pm$0.02 & 16.54$\pm$0.04 \\
  04:06:21.18 & +47:47:57.1 & 114$\pm$8 & -199$\pm$8 & 19.12$\pm$0.02 & 17.94$\pm$0.03 & 16.38$\pm$0.02 & 15.79$\pm$0.02 & 15.51$\pm$0.02 \\
  04:06:33.73 & +47:15:16.7 & 71$\pm$7 & -220$\pm$7 & 21.09$\pm$0.14 & 17.94$\pm$0.03 &      &      & 13.80$\pm$0.02 \\
  04:07:29.52 & +58:38:55.9 & 81$\pm$7 & -318$\pm$7 & 19.61$\pm$0.03 & 16.57$\pm$0.02 &      &      & 13.02$\pm$0.02 \\
  04:08:23.65 & +47:05:03.0 & 88$\pm$8 & -337$\pm$6 & 21.30$\pm$0.14 & 18.12$\pm$0.03 & 14.68$\pm$0.02 & 14.27$\pm$0.02 & 13.77$\pm$0.02 \\
  04:13:22.06 & +44:44:36.6 & 162$\pm$7 & -183$\pm$4 & 20.75$\pm$0.27 & 17.66$\pm$0.03 & 14.69$\pm$0.02 & 14.15$\pm$0.02 & 13.67$\pm$0.02 \\
  04:22:28.37 & +45:17:39.8 & 273$\pm$7 & -13$\pm$6 &      & 17.80$\pm$0.03 & 15.53$\pm$0.02 & 15.14$\pm$0.02 & 14.78$\pm$0.02 \\
  04:30:46.79 & +49:55:49.7 & 157$\pm$8 & -170$\pm$9 & 19.92$\pm$0.04 & 17.47$\pm$0.02 & 15.37$\pm$0.02 & 14.95$\pm$0.02 & 14.62$\pm$0.02 \\
  04:31:21.84 & +47:07:38.3 & 403$\pm$7 & -121$\pm$9 & 19.06$\pm$0.04 & 17.45$\pm$0.03 & 15.72$\pm$0.02 & 15.16$\pm$0.02 & 14.89$\pm$0.02 \\
  04:34:03.07 & +49:53:40.0 & 91$\pm$8 & -183$\pm$8 & 19.10$\pm$0.03 & 16.61$\pm$0.02 &      &      & 13.50$\pm$0.02 \\
  04:35:19.94 & +43:06:09.4 & 148$\pm$4 & -168$\pm$5 & 15.86$\pm$0.02 & 13.92$\pm$0.02 & 11.98$\pm$0.02 & 11.44$\pm$0.02 & 11.13$\pm$0.02 \\
  04:43:30.77 & +46:45:34.4 & 236$\pm$6 & -105$\pm$5 &      & 18.7$\pm$0.1$^{ae}$ & 15.85$\pm$0.02 & 15.44$\pm$0.02 & 15.00$\pm$0.02 \\ 
  04:44:21.20 & +42:02:36.2 & 110$\pm$4 & -170$\pm$4 & 18.92$\pm$0.04 & 16.70$\pm$0.02 &      &      & 13.91$\pm$0.02 \\
  04:46:39.22 & +40:54:52.9 & 191$\pm$4 & -162$\pm$5 &      & 20.9$\pm$0.5$^a$ & 16.84$\pm$0.02 & 16.21$\pm$0.02 & 15.60$\pm$0.02 \\ 
  04:51:43.83 & +45:49:58.0 & 365$\pm$5 & -574$\pm$5 &      & 13.22$\pm$0.02 & 11.7$\pm$0.2$^b$ & 11.4$\pm$0.2$^b$ & 10.7$\pm$0.2$^b$ \\ 
  04:54:09.74 & +46:16:09.4 & 195$\pm$13 & -133$\pm$13 &      & 19.48$\pm$0.11 & 17.38$\pm$0.02 & 16.94$\pm$0.02 & 16.59$\pm$0.04 \\
  05:07:19.98 & +34:58:52.6 & 60$\pm$3 & -277$\pm$4 & 19.45$\pm$0.03 & 16.99$\pm$0.02 & 14.68$\pm$0.02 & 14.23$\pm$0.02 & 13.83$\pm$0.02 \\
  05:15:37.00 & +44:48:20.1 & 58$\pm$5 & -280$\pm$5 & 16.93$\pm$0.02 & 15.83$\pm$0.02 & 14.60$\pm$0.02 & 14.13$\pm$0.02 & 13.90$\pm$0.02 \\
  05:18:17.90 & +42:11:31.9 & -15$\pm$5 & -238$\pm$3 &      &      & 14.89$\pm$0.02 & 14.31$\pm$0.02 & 13.80$\pm$0.02 \\
  05:21:24.85 & +39:08:14.1 & 121$\pm$4 & -237$\pm$4 & 18.50$\pm$0.02 & 17.60$\pm$0.03 & 16.93$\pm$0.02 & 16.45$\pm$0.02 & 16.26$\pm$0.03 \\
  05:22:49.68 & +43:51:57.7 & 103$\pm$4 & -208$\pm$5 & 16.58$\pm$0.02 & 15.09$\pm$0.02 & 13.53$\pm$0.02 & 13.05$\pm$0.02 & 12.77$\pm$0.02 \\
  05:22:50.00 & +39:41:28.6 & 205$\pm$3 & -119$\pm$3 & 14.54$\pm$0.02 & 13.31$\pm$0.02 & 11.97$\pm$0.02 & 11.47$\pm$0.02 & 11.21$\pm$0.02 \\
  05:25:55.61 & +37:22:11.9 & 129$\pm$7 & -175$\pm$7 &      & 19.48$\pm$0.10 & 17.43$\pm$0.03 & 17.01$\pm$0.04 & 16.66$\pm$0.05 \\
  05:25:59.58 & +33:06:06.3 & 85$\pm$6 & -193$\pm$7 & 20.17$\pm$0.08 & 17.52$\pm$0.02 & 15.19$\pm$0.02 & 14.74$\pm$0.02 & 14.33$\pm$0.02 \\
  05:27:21.68 & +42:58:36.5 & 155$\pm$5 & -304$\pm$5 & 16.59$\pm$0.02 & 14.94$\pm$0.02 & 13.34$\pm$0.02 & 12.89$\pm$0.02 & 12.61$\pm$0.02 \\
  05:33:01.65 & +37:38:20.0 & 44$\pm$4 & -245$\pm$4 &      & $>$20.5$^d$ & 16.79$\pm$0.02 & 15.94$\pm$0.02 & 15.16$\pm$0.02 \\ 
  05:34:49.33 & +40:37:01.4 & -40$\pm$4 & -277$\pm$4 & 20.84$\pm$0.08 & 18.28$\pm$0.04 & 16.05$\pm$0.02 & 15.63$\pm$0.02 & 15.32$\pm$0.02 \\
  05:35:42.81 & +26:49:20.4 & 173$\pm$7 & -364$\pm$8 &      & 18.49$\pm$0.05 & 15.77$\pm$0.02 & 15.35$\pm$0.02 & 14.96$\pm$0.02 \\
  05:40:41.35 & +34:00:24.1 & 142$\pm$6 & -172$\pm$6 & 16.07$\pm$0.02 & 15.09$\pm$0.02 & 13.85$\pm$0.02 & 13.28$\pm$0.02 & 13.05$\pm$0.02 \\
  05:41:47.12 & +30:20:16.2 & 177$\pm$6 & -97$\pm$6 & 17.42$\pm$0.02 & 16.09$\pm$0.02 & 14.63$\pm$0.02 & 14.07$\pm$0.02 & 13.80$\pm$0.02 \\
  05:44:14.33 & +34:55:42.9 & 64$\pm$4 & -205$\pm$4 & 21.31$\pm$0.13 & 18.56$\pm$0.04 &      &      & 14.76$\pm$0.02 \\
  05:44:57.43 & +37:05:04.1 & -10$\pm$3 & -228$\pm$4 & 21.69$\pm$0.41 & 18.01$\pm$0.04 & 13.90$\pm$0.02 & 13.10$\pm$0.02 & 12.36$\pm$0.02 \\
  05:45:02.32 & +33:58:31.6 & 66$\pm$6 & -206$\pm$6 & 20.62$\pm$0.09 & 17.56$\pm$0.02 & 15.11$\pm$0.02 & 14.68$\pm$0.02 & 14.29$\pm$0.02 \\
  05:47:25.11 & +35:42:43.0 & 134$\pm$4 & -190$\pm$3 & 19.05$\pm$0.03 & 17.72$\pm$0.03 & 16.30$\pm$0.02 & 15.83$\pm$0.02 & 15.56$\pm$0.02 \\
  05:47:33.08 & +38:03:05.5 & 198$\pm$5 & -181$\pm$4 & 17.72$\pm$0.02 & 16.79$\pm$0.02 & 15.73$\pm$0.02 & 15.20$\pm$0.02 & 15.02$\pm$0.02 \\
  05:49:02.26 & +22:23:30.5 & 10$\pm$9 & -249$\pm$9 & 19.51$\pm$0.03 & 16.57$\pm$0.02 & 13.43$\pm$0.02 & 12.81$\pm$0.02 & 12.34$\pm$0.02 \\
  05:54:03.58 & +20:28:36.8 & 209$\pm$6 & -221$\pm$7 & 20.17$\pm$0.20$^e$ & 18.23$\pm$0.20$^e$ & 15.85$\pm$0.02 & 15.44$\pm$0.02 & 15.15$\pm$0.02 \\ 
  05:54:07.07 & +21:18:20.0 & 126$\pm$11 & -189$\pm$11 & 20.21$\pm$0.06 & 18.80$\pm$0.06 & 17.25$\pm$0.02 & 16.81$\pm$0.03 & 16.56$\pm$0.04 \\
  05:54:46.14 & +19:31:13.6 & 174$\pm$8 & -168$\pm$8 & 19.39$\pm$0.04 & 17.68$\pm$0.02 & 16.11$\pm$0.02 & 15.65$\pm$0.02 & 15.35$\pm$0.02 \\
  05:55:39.76 & +22:52:55.3 & 18$\pm$10 & -243$\pm$9 & 20.42$\pm$0.07 & 18.64$\pm$0.06 & 16.80$\pm$0.02 & 16.31$\pm$0.02 & 16.01$\pm$0.02 \\
  05:57:17.90 & +34:37:14.1 & -69$\pm$3 & -197$\pm$3 &      & 19.34$\pm$0.09 & 16.09$\pm$0.02 & 15.66$\pm$0.02 & 15.20$\pm$0.02 \\
  06:02:31.77 & +22:15:53.3 & 19$\pm$13 & -280$\pm$13 &      & $>$20.6$^d$ & 17.82$\pm$0.03 & 17.29$\pm$0.04 & 16.67$\pm$0.05 \\ 
  06:03:45.45 & +20:16:44.8 & 39$\pm$10 & -200$\pm$10 &      & 19.9$\pm$0.3$^a$ & 17.29$\pm$0.02 & 16.82$\pm$0.02 & 16.47$\pm$0.04 \\ 
  06:05:18.50 & +19:48:49.9 & 60$\pm$5 & -216$\pm$6 & 18.46$\pm$0.02 & 16.02$\pm$0.02 & 13.80$\pm$0.02 & 13.37$\pm$0.02 & 13.01$\pm$0.02 \\
  06:10:37.22 & +12:58:48.6 & 14$\pm$9 & -228$\pm$6 & 20.82$\pm$0.15 & 17.69$\pm$0.03 & 14.88$\pm$0.02 & 14.38$\pm$0.02 & 13.97$\pm$0.02 \\
  06:21:09.24 & +08:23:15.6 & 164$\pm$7 & -115$\pm$8 & 19.19$\pm$0.02 & 17.29$\pm$0.02 & 15.42$\pm$0.02 & 14.99$\pm$0.02 & 14.73$\pm$0.02 \\
  06:24:40.48 & +06:48:22.3 & 94$\pm$5 & -197$\pm$5 & 19.48$\pm$0.03 & 18.26$\pm$0.04 & 16.78$\pm$0.02 & 16.34$\pm$0.02 & 16.04$\pm$0.03 \\
  06:27:27.79 & +05:15:37.5 & 119$\pm$6 & -319$\pm$6 & 20.75$\pm$0.32 & 18.16$\pm$0.04 & 16.22$\pm$0.02 & 15.81$\pm$0.02 & 15.51$\pm$0.02 \\
  06:27:50.06 & +01:43:29.6 & 213$\pm$4 & -154$\pm$5 &      & $>$20.4$^d$ & 16.89$\pm$0.02 & 16.30$\pm$0.02 & 15.66$\pm$0.02 \\ 
  06:30:50.26 & -00:10:49.7 & -43$\pm$4 & -291$\pm$4 & 18.48$\pm$0.04 & 16.56$\pm$0.02 & 14.71$\pm$0.02 & 14.25$\pm$0.02 & 13.97$\pm$0.02 \\
  06:32:32.21 & +01:53:49.8 & 55$\pm$4 & -216$\pm$4 & 18.85$\pm$0.05 & 17.59$\pm$0.03 & 16.13$\pm$0.02 & 15.69$\pm$0.02 & 15.45$\pm$0.02 \\
  06:32:35.71 & +03:14:03.7 & -167$\pm$5 & 166$\pm$6 &      & 17.93$\pm$0.03 & 14.61$\pm$0.02 & 14.06$\pm$0.02 & 13.58$\pm$0.02 \\
  \hline
  \multicolumn{9}{|l|}{$^a$ Based on our own aperture photometry due to the source being a non-detection in IPHAS i band catalogue but visible in image.}\\
\multicolumn{9}{|l|}{$^b$ Based on our own aperture photometry, survey photometry was deemed unreliable due to nearby bright companion.}\\
\multicolumn{9}{|l|}{$^c$ Undetected in IPHAS source identification. Profile fit photometry performed to measure this magnitude.}\\
\multicolumn{9}{|l|}{$^d$ Undetected in IPHAS i band, 3$\sigma$ limit of field given.}\\
\multicolumn{9}{|l|}{$^e$ Based on IPHAS observations in poor weather, this photometry is unreliable.}\\
  \end{tabular}
\end{minipage}
\end{table*}

\newpage

\begin{table*}
\begin{minipage}{155mm}
\contcaption{}
\begin{tabular}{|l|l|l|l|l|l|l|l|l|}
\hline
  \multicolumn{1}{|c|}{$\alpha$} &
  \multicolumn{1}{c|}{$\delta$} &
  \multicolumn{1}{c|}{$\mu_{\alpha} \cos \delta$} &
  \multicolumn{1}{c|}{$\mu_{\delta}$} &
  \multicolumn{1}{c|}{r} &
  \multicolumn{1}{c|}{i} &
  \multicolumn{1}{c|}{J} &
  \multicolumn{1}{c|}{H} &
  \multicolumn{1}{c|}{K} \\
\hline
  06:33:33.37 & +10:01:27.4 & 184$\pm$15 & -214$\pm$12 & 19.28$\pm$0.20$^e$ & 19.21$\pm$0.20$^e$ & 15.15$\pm$0.02 & 14.72$\pm$0.02 & 14.21$\pm$0.02 \\ 
  06:43:16.71 & -06:10:10.3 & 239$\pm$4 & -109$\pm$4 &      &      & 14.96$\pm$0.02 & 14.52$\pm$0.02 & 14.23$\pm$0.02 \\
  06:44:16.75 & +03:16:17.9 & 101$\pm$3 & -280$\pm$3 &      & 17.3$\pm$0.2$^{ae}$ & 14.06$\pm$0.02 & 13.47$\pm$0.02 & 12.99$\pm$0.02 \\ 
  06:45:04.23 & -02:52:49.3 & 149$\pm$5 & -231$\pm$4 & 18.55$\pm$0.02 & 17.37$\pm$0.03 & 16.00$\pm$0.02 & 15.50$\pm$0.02 & 15.22$\pm$0.02 \\
  06:45:33.53 & -05:30:47.2 & 58$\pm$4 & -218$\pm$3 &      &      & 14.28$\pm$0.02 & 13.72$\pm$0.02 & 13.37$\pm$0.02 \\
  06:45:49.25 & -05:22:19.9 & 297$\pm$4 & -61$\pm$5 &      &      & 16.44$\pm$0.02 & 15.90$\pm$0.02 & 15.70$\pm$0.02 \\
  06:47:12.29 & +08:24:40.0 & 150$\pm$5 & -195$\pm$6 &      & 20.2$\pm$0.2$^a$ & 16.39$\pm$0.02 & 15.76$\pm$0.02 & 15.17$\pm$0.02 \\ 
  06:47:24.04 & -03:38:38.7 & 63$\pm$4 & -203$\pm$4 &      &      & 15.85$\pm$0.02 & 15.41$\pm$0.02 & 15.16$\pm$0.02 \\
  06:50:05.51 & -03:03:32.8 & 34$\pm$3 & -270$\pm$3 & 17.49$\pm$0.20$^e$ & 15.45$\pm$0.20$^e$ & 13.10$\pm$0.02 & 12.52$\pm$0.02 & 12.24$\pm$0.02 \\ 
  06:52:10.42 & -01:20:35.5 & 194$\pm$6 & -233$\pm$9 & 18.49$\pm$0.02 & 17.55$\pm$0.02 & 16.17$\pm$0.02 & 15.64$\pm$0.02 & 15.38$\pm$0.02 \\
  06:53:27.95 & +02:39:43.8 & -15$\pm$3 & 210$\pm$4 & 16.91$\pm$0.02 & 16.58$\pm$0.02 & 16.24$\pm$0.02 & 16.00$\pm$0.02 & 15.94$\pm$0.02 \\
  06:54:12.55 & -06:27:31.0 & 260$\pm$4 & -271$\pm$3 &      &      &      &      & 14.99$\pm$0.02 \\
  06:55:37.37 & -06:38:04.2 & 66$\pm$3 & -281$\pm$4 &      &      & 14.01$\pm$0.02 & 13.48$\pm$0.02 & 13.11$\pm$0.02 \\
  06:56:17.02 & +04:21:05.7 & 35$\pm$4 & -228$\pm$4 & 19.35$\pm$0.02 & 17.61$\pm$0.02 & 15.97$\pm$0.02 & 15.51$\pm$0.02 & 15.22$\pm$0.02 \\
  06:58:45.22 & -01:15:52.6 & 271$\pm$9 & -293$\pm$8 & 17.72$\pm$0.02 & 17.15$\pm$0.02 & 16.57$\pm$0.02 & 16.32$\pm$0.02 & 16.41$\pm$0.06 \\
  07:01:52.79 & -13:02:31.2 & 39$\pm$5 & -230$\pm$4 & 19.47$\pm$0.04 & 18.27$\pm$0.02 & 16.68$\pm$0.02 & 16.21$\pm$0.02 & 15.98$\pm$0.03 \\
  07:05:18.20 & +01:29:10.4 & 167$\pm$3 & -268$\pm$3 &      & 17.85$\pm$0.02 & 14.99$\pm$0.02 & 14.48$\pm$0.02 & 14.07$\pm$0.02 \\
  07:06:52.01 & -17:05:19.1 & -334$\pm$4 & -71$\pm$4 &      &      & 12.75$\pm$0.02 & 12.37$\pm$0.02 & 12.00$\pm$0.02 \\
  07:07:29.26 & -02:49:48.6 & 61$\pm$3 & 194$\pm$4 &      &      & 12.06$\pm$0.02 & 11.52$\pm$0.02 & 11.29$\pm$0.02 \\
  07:09:35.53 & -15:08:08.1 & -113$\pm$4 & -294$\pm$4 &      &      & 13.38$\pm$0.02 & 12.95$\pm$0.02 & 12.67$\pm$0.02 \\
  07:09:39.72 & -12:38:46.4 & -47$\pm$6 & -267$\pm$5 & 16.60$\pm$0.02 & 15.12$\pm$0.02 & 13.53$\pm$0.02 & 13.08$\pm$0.02 & 12.82$\pm$0.02 \\
  07:12:18.30 & -07:59:33.7 & 61$\pm$3 & -213$\pm$3 &      &      & 13.35$\pm$0.02 & 12.79$\pm$0.02 & 12.64$\pm$0.02 \\
  07:12:44.28 & -12:58:02.4 & 151$\pm$5 & -309$\pm$5 & 19.48$\pm$0.03 & 17.72$\pm$0.02 & 15.64$\pm$0.02 & 15.15$\pm$0.02 & 14.84$\pm$0.02 \\
  07:17:12.18 & -14:55:13.8 & -16$\pm$6 & -244$\pm$8 &      &      & 13.48$\pm$0.02 & 13.02$\pm$0.02 & 12.74$\pm$0.02 \\
  07:17:28.15 & -15:05:17.6 & 62$\pm$4 & -223$\pm$5 & 13.40$\pm$0.02 & 12.85$\pm$0.02 & 11.93$\pm$0.02 & 11.35$\pm$0.02 & 11.35$\pm$0.02 \\
  07:22:22.47 & -14:51:22.4 & -347$\pm$6 & 133$\pm$4 & 15.95$\pm$0.02 & 13.92$\pm$0.02 & 11.77$\pm$0.02 & 11.79$\pm$0.02 & 11.00$\pm$0.02 \\
  07:24:15.02 & -12:47:31.2 & 174$\pm$5 & -228$\pm$5 & 19.16$\pm$0.02 & 17.90$\pm$0.02 & 16.37$\pm$0.02 & 15.90$\pm$0.02 & 15.68$\pm$0.02 \\
  07:28:53.82 & -14:01:10.2 & -270$\pm$4 & -67$\pm$3 &      &      & 13.95$\pm$0.02 & 13.51$\pm$0.02 & 13.17$\pm$0.02 \\
  07:29:45.63 & -07:33:49.1 & 208$\pm$4 & -114$\pm$4 &      &      &      & 14.57$\pm$0.02 & 14.21$\pm$0.02 \\
  07:40:28.99 & -13:00:44.1 & 31$\pm$4 & 221$\pm$5 &      &      & 11.77$\pm$0.02 & 10.74$\pm$0.02 & 11.05$\pm$0.02 \\
  19:55:30.48 & +26:13:13.1 & -185$\pm$10 & -96$\pm$10 &      & $>$20.7$^d$ & 18.48$\pm$0.05 & 17.66$\pm$0.04 & 16.99$\pm$0.07 \\ 
  19:59:24.81 & +39:31:56.1 & -62$\pm$9 & 321$\pm$7 & 13.74$\pm$0.02 & 12.75$\pm$0.02 & 11.02$\pm$0.02 & 10.17$\pm$0.02 & 11.37$\pm$0.02 \\
  20:04:09.96 & +24:26:35.3 & -109$\pm$5 & -168$\pm$5 &      & 20.04$\pm$0.22 & 16.90$\pm$0.02 & 16.48$\pm$0.02 & 16.09$\pm$0.03 \\
  20:05:30.35 & +36:35:50.7 & -126$\pm$6 & -184$\pm$6 &      &      & 17.47$\pm$0.02 & 16.61$\pm$0.02 & 16.01$\pm$0.04 \\
  20:05:35.97 & +30:09:24.8 & 96$\pm$3 & -216$\pm$3 & 19.45$\pm$0.04 & 16.50$\pm$0.02 & 13.92$\pm$0.02 & 13.53$\pm$0.02 & 13.13$\pm$0.02 \\
  20:06:11.93 & +24:07:53.3 & 226$\pm$3 & 226$\pm$3 &      & 16.98$\pm$0.02 & 13.96$\pm$0.02 & 13.42$\pm$0.02 & 13.00$\pm$0.02 \\
  20:08:37.94 & +39:20:56.1 & -45$\pm$11 & -199$\pm$12 & 19.85$\pm$0.06 & 17.39$\pm$0.02 & 15.17$\pm$0.02 & 14.69$\pm$0.02 & 14.34$\pm$0.02 \\
  20:09:12.58 & +37:12:12.5 & -137$\pm$4 & -199$\pm$5 & 24.63$\pm$3.30 & 19.50$\pm$0.11 & 15.73$\pm$0.02 & 15.11$\pm$0.02 & 14.58$\pm$0.02 \\
  20:14:39.96 & +38:50:23.4 & -218$\pm$6 & -249$\pm$6 & 16.09$\pm$0.02 & 14.16$\pm$0.02 & 12.23$\pm$0.02 & 11.71$\pm$0.02 & 11.54$\pm$0.02 \\
  20:16:08.23 & +42:00:32.8 & 237$\pm$6 & 152$\pm$6 & 18.67$\pm$0.02 & 16.50$\pm$0.02 & 14.46$\pm$0.02 & 14.11$\pm$0.02 & 13.75$\pm$0.02 \\
  20:16:44.23 & +41:40:43.5 & -23$\pm$7 & -211$\pm$7 & 16.21$\pm$0.02 & 15.39$\pm$0.02 & 14.27$\pm$0.02 & 13.76$\pm$0.02 & 13.59$\pm$0.02 \\
  20:17:31.27 & +33:53:59.5 & -247$\pm$4 & 24$\pm$4 &      & $>$20.4$^d$ & 17.35$\pm$0.02 & 16.61$\pm$0.03 & 16.01$\pm$0.03 \\ 
  20:19:02.41 & +44:53:16.3 & -173$\pm$6 & -140$\pm$5 & 17.21$\pm$0.02 & 15.35$\pm$0.02 & 13.51$\pm$0.02 & 13.06$\pm$0.02 & 12.80$\pm$0.02 \\
  20:21:12.34 & +37:25:05.9 & -171$\pm$3 & -234$\pm$3 & 19.69$\pm$0.05 & 16.64$\pm$0.02 & 13.98$\pm$0.02 & 13.46$\pm$0.02 & 13.10$\pm$0.02 \\
  20:25:28.30 & +40:13:19.7 & 184$\pm$6 & 106$\pm$7 & 16.74$\pm$0.02 & 15.07$\pm$0.02 & 13.42$\pm$0.02 & 12.95$\pm$0.02 & 12.68$\pm$0.02 \\
  20:28:02.85 & +42:05:51.6 & -169$\pm$23 & -189$\pm$20 &      &      &      &      & 16.47$\pm$0.06 \\
  20:28:10.47 & +41:01:19.0 & 177$\pm$4 & 123$\pm$4 & 18.90$\pm$0.02 & 16.37$\pm$0.02 & 14.00$\pm$0.02 & 13.51$\pm$0.02 & 13.13$\pm$0.02 \\
  20:31:32.54 & +43:03:19.4 & 259$\pm$5 & 326$\pm$7 &      & 20.29$\pm$0.17 & 15.85$\pm$0.02 & 15.02$\pm$0.02 & 14.27$\pm$0.02 \\
  20:32:53.25 & +36:38:17.0 & 28$\pm$7 & 200$\pm$7 &      & 19.32$\pm$0.12 & 17.03$\pm$0.02 & 16.67$\pm$0.02 & 16.37$\pm$0.04 \\
  20:33:46.52 & +37:50:58.3 & 197$\pm$4 & 49$\pm$5 & 18.35$\pm$0.02 & 16.23$\pm$0.02 & 14.11$\pm$0.02 & 13.59$\pm$0.02 & 13.26$\pm$0.02 \\
  20:43:14.92 & +50:46:22.5 & 170$\pm$7 & 199$\pm$7 &      &      & 15.12$\pm$0.02 & 14.65$\pm$0.02 & 14.24$\pm$0.02 \\
  20:48:00.13 & +50:38:19.3 & 99$\pm$8 & 248$\pm$8 &      &      & 16.30$\pm$0.02 & 16.37$\pm$0.02 & 16.32$\pm$0.03 \\
  20:53:10.56 & +36:27:37.6 & 95$\pm$4 & 219$\pm$4 &      &      & 12.29$\pm$0.02 & 11.33$\pm$0.02 & 11.46$\pm$0.02 \\
  20:55:07.49 & +38:46:44.4 & 139$\pm$5 & 155$\pm$5 & 19.81$\pm$0.07 & 17.83$\pm$0.02 & 16.02$\pm$0.02 & 15.56$\pm$0.02 & 15.28$\pm$0.02 \\
  20:55:25.17 & +45:06:10.4 & 40$\pm$3 & -236$\pm$3 & 18.16$\pm$0.02 & 15.97$\pm$0.02 & 13.83$\pm$0.02 & 13.37$\pm$0.02 & 13.02$\pm$0.02 \\
  20:55:27.42 & +38:12:17.1 & 176$\pm$4 & 193$\pm$5 & 18.53$\pm$0.04 & 16.76$\pm$0.02 & 15.00$\pm$0.02 & 14.55$\pm$0.02 & 14.27$\pm$0.02 \\
  20:59:58.64 & +41:25:13.7 & -78$\pm$7 & 193$\pm$6 &      & $>$20.5$^d$ & 16.91$\pm$0.02 & 16.38$\pm$0.02 & 15.77$\pm$0.02 \\ 
  21:04:53.81 & +41:46:28.3 & 154$\pm$4 & 141$\pm$4 & 17.50$\pm$0.02 & 15.40$\pm$0.02 & 13.43$\pm$0.02 & 12.98$\pm$0.02 & 12.65$\pm$0.02 \\
  \hline\end{tabular}
\end{minipage}
\end{table*}

\newpage

\begin{table*}
\begin{minipage}{155mm}
\contcaption{}
\begin{tabular}{|l|l|l|l|l|l|l|l|l|}
\hline
  \multicolumn{1}{|c|}{$\alpha$} &
  \multicolumn{1}{c|}{$\delta$} &
  \multicolumn{1}{c|}{$\mu_{\alpha} \cos \delta$} &
  \multicolumn{1}{c|}{$\mu_{\delta}$} &
  \multicolumn{1}{c|}{r} &
  \multicolumn{1}{c|}{i} &
  \multicolumn{1}{c|}{J} &
  \multicolumn{1}{c|}{H} &
  \multicolumn{1}{c|}{K} \\
\hline
  21:05:57.74 & +47:01:44.5 & 126$\pm$4 & 174$\pm$3 & 14.82$\pm$0.02 & 13.37$\pm$0.02 & 11.84$\pm$0.02 & 11.20$\pm$0.02 & 11.07$\pm$0.02 \\
  21:06:04.96 & +50:19:54.9 & -140$\pm$7 & -174$\pm$7 &      & 19.69$\pm$0.12 & 15.95$\pm$0.02 & 15.52$\pm$0.02 & 15.06$\pm$0.02 \\
  21:07:35.11 & +48:13:43.6 & 170$\pm$6 & 181$\pm$7 & 19.76$\pm$0.12 & 16.80$\pm$0.02 & 14.12$\pm$0.02 & 13.62$\pm$0.02 & 13.22$\pm$0.02 \\
  21:09:29.87 & +50:24:34.5 & 132$\pm$5 & -158$\pm$6 & 19.80$\pm$0.07 & 18.29$\pm$0.05 & 16.83$\pm$0.02 & 16.41$\pm$0.02 & 16.21$\pm$0.03 \\
  21:11:04.39 & +48:00:21.9 & 193$\pm$3 & 127$\pm$3 &      & 14.60$\pm$0.02 & 12.72$\pm$0.02 & 12.21$\pm$0.02 & 11.87$\pm$0.02 \\
  21:11:54.34 & +43:17:54.1 & 143$\pm$3 & 140$\pm$3 & 19.85$\pm$0.07 & 17.53$\pm$0.02 & 15.26$\pm$0.02 & 14.77$\pm$0.02 & 14.45$\pm$0.02 \\
  21:14:25.17 & +50:10:15.9 & 296$\pm$5 & 30$\pm$4 &      & 20.00$\pm$0.19 & 16.08$\pm$0.02 & 15.30$\pm$0.02 & 14.51$\pm$0.02 \\
  21:18:59.25 & +43:38:01.3 & 169$\pm$3 & 134$\pm$3 &      & 17.8$\pm$0.2$^c$ & 14.7$\pm$0.2$^b$ & 14.1$\pm$0.2$^b$ & 13.8$\pm$0.2$^b$ \\ 
  21:19:52.83 & +48:26:12.0 & 151$\pm$4 & 216$\pm$3 &      & 20.37$\pm$0.26 & 15.98$\pm$0.02 & 15.32$\pm$0.02 & 14.74$\pm$0.02 \\
  21:20:11.58 & +51:14:15.9 & 46$\pm$4 & 220$\pm$4 & 18.10$\pm$0.02 & 16.46$\pm$0.02 & 14.85$\pm$0.02 & 14.42$\pm$0.02 & 14.12$\pm$0.02 \\
  21:28:05.25 & +56:34:16.6 & 155$\pm$8 & 135$\pm$8 & 18.95$\pm$0.02 & 17.61$\pm$0.02 & 16.16$\pm$0.02 & 15.65$\pm$0.02 & 15.41$\pm$0.02 \\
  21:29:19.99 & +51:46:42.4 & -92$\pm$4 & 214$\pm$4 & 21.59$\pm$0.17 & 18.37$\pm$0.04 & 17.47$\pm$0.02 & 17.27$\pm$0.04 & 14.88$\pm$0.02 \\
  21:30:48.11 & +57:12:36.8 & 298$\pm$4 & 91$\pm$6 & 21.51$\pm$0.21 & 17.93$\pm$0.03 & 14.90$\pm$0.02 & 14.42$\pm$0.02 & 13.98$\pm$0.02 \\
  21:32:12.97 & +44:52:29.3 & 172$\pm$6 & 145$\pm$7 & 19.61$\pm$0.06 & 17.97$\pm$0.03 & 16.39$\pm$0.02 & 15.92$\pm$0.02 & 15.67$\pm$0.02 \\
  21:41:15.07 & +56:40:12.9 & 234$\pm$8 & 246$\pm$9 & 21.28$\pm$0.15 & 18.07$\pm$0.03 & 15.05$\pm$0.02 & 14.48$\pm$0.02 & 14.01$\pm$0.02 \\
  21:44:46.49 & +50:52:26.6 & -99$\pm$5 & -203$\pm$4 &      & 18.62$\pm$0.05 & 15.62$\pm$0.02 & 15.12$\pm$0.02 & 14.71$\pm$0.02 \\
  21:52:50.45 & +55:12:52.9 & 228$\pm$4 & 28$\pm$4 & 17.57$\pm$0.02 & 15.60$\pm$0.02 & 13.69$\pm$0.02 & 13.25$\pm$0.02 & 12.93$\pm$0.02 \\
  21:53:12.31 & +53:51:49.6 & 243$\pm$4 & 30$\pm$4 & 19.67$\pm$0.07 & 17.33$\pm$0.02 & 15.19$\pm$0.02 & 14.76$\pm$0.02 & 14.44$\pm$0.02 \\
  22:00:15.74 & +51:09:45.5 & 123$\pm$13 & 182$\pm$13 &      &      & 17.78$\pm$0.03 & 17.38$\pm$0.04 & 17.10$\pm$0.07 \\
  22:00:44.34 & +54:21:55.6 & 218$\pm$4 & 126$\pm$4 &      & 20.07$\pm$0.24 & 16.32$\pm$0.02 & 15.71$\pm$0.02 & 15.20$\pm$0.02 \\
  22:01:56.74 & +51:08:19.7 & 206$\pm$6 & 74$\pm$6 & 17.30$\pm$0.02 & 14.80$\pm$0.02 & 12.35$\pm$0.02 & 11.90$\pm$0.02 & 11.47$\pm$0.02 \\
  22:07:06.47 & +53:52:53.3 & 154$\pm$4 & 215$\pm$3 &      & 20.31$\pm$0.22 & 16.22$\pm$0.02 & 15.59$\pm$0.02 & 15.00$\pm$0.02 \\
  22:07:15.68 & +52:14:05.1 & 103$\pm$4 & 225$\pm$4 & 18.19$\pm$0.02 & 17.01$\pm$0.02 & 15.54$\pm$0.02 & 15.10$\pm$0.02 & 14.88$\pm$0.02 \\
  22:10:32.97 & +52:55:51.2 & 206$\pm$6 & 112$\pm$7 & 13.69$\pm$0.02 & 11.82$\pm$0.02 & 11.89$\pm$0.02 & 11.28$\pm$0.02 & 11.06$\pm$0.02 \\
  22:11:49.28 & +55:02:06.2 & -121$\pm$3 & -237$\pm$3 & 17.36$\pm$0.02 & 15.35$\pm$0.02 & 13.35$\pm$0.02 & 12.93$\pm$0.02 & 12.63$\pm$0.02 \\
  22:13:07.90 & +52:55:08.7 & -201$\pm$6 & -38$\pm$6 & 18.65$\pm$0.02 & 16.27$\pm$0.02 & 14.02$\pm$0.02 & 13.49$\pm$0.02 & 13.16$\pm$0.02 \\
  22:17:30.73 & +51:39:32.5 & 217$\pm$5 & 32$\pm$4 & 21.36$\pm$0.16 & 18.06$\pm$0.03 & 15.11$\pm$0.02 & 14.56$\pm$0.02 & 14.12$\pm$0.02 \\
  22:18:00.59 & +56:02:15.1 & 108$\pm$4 & 227$\pm$4 & 17.88$\pm$0.02 & 17.21$\pm$0.02 & 16.58$\pm$0.02 & 16.40$\pm$0.02 & 16.42$\pm$0.04 \\
  22:18:17.08 & +56:12:28.9 & 236$\pm$3 & 21$\pm$3 & 19.84$\pm$0.04 & 17.29$\pm$0.02 & 15.12$\pm$0.02 & 14.71$\pm$0.02 & 14.37$\pm$0.02 \\
  22:21:29.14 & +55:56:00.1 & 106$\pm$5 & 176$\pm$4 & 18.47$\pm$0.02 & 16.68$\pm$0.02 & 14.95$\pm$0.02 & 14.48$\pm$0.02 & 14.20$\pm$0.02 \\
  22:21:35.00 & +54:09:52.5 & 215$\pm$3 & 176$\pm$4 & 17.45$\pm$0.02 & 15.67$\pm$0.02 & 13.81$\pm$0.02 & 13.40$\pm$0.02 & 13.12$\pm$0.02 \\
  22:23:04.16 & +54:28:00.8 & 166$\pm$6 & 124$\pm$6 & 20.13$\pm$0.12 & 18.78$\pm$0.10 & 17.36$\pm$0.02 & 16.92$\pm$0.03 & 16.66$\pm$0.05 \\
  22:32:30.30 & +55:15:24.3 & 202$\pm$5 & 69$\pm$6 & 18.73$\pm$0.03 & 17.78$\pm$0.03 & 16.35$\pm$0.02 & 15.78$\pm$0.02 & 15.54$\pm$0.02 \\
  22:32:33.83 & +52:35:22.5 & 229$\pm$5 & 38$\pm$5 & 16.34$\pm$0.02 & 14.49$\pm$0.02 & 12.77$\pm$0.02 & 12.34$\pm$0.02 & 12.05$\pm$0.02 \\
  22:33:16.60 & +52:27:22.3 & 212$\pm$5 & 8$\pm$6 & 16.58$\pm$0.02 & 14.26$\pm$0.02 & 12.06$\pm$0.02 & 11.59$\pm$0.02 & 11.18$\pm$0.02 \\
\hline
\end{tabular}
\end{minipage}
\end{table*}

\clearpage
	\begin{table*}
\begin{minipage}{120mm}
\caption{UKIDSS GPS proper motions and epoch 2000.0 coordinates for twelve recent WISE discoveries.}\label{WISEPMs}
\begin{tabular}{|l|l|l|l|l|l|}
\hline
  \multicolumn{1}{|c|}{$\alpha$} &
  \multicolumn{1}{c|}{$\delta$} &
  \multicolumn{1}{c|}{$\mu_{\alpha} \cos \delta$} &
  \multicolumn{1}{c|}{$\mu_{\delta}$} &
  \multicolumn{1}{c|}{K} &
  \multicolumn{1}{c|}{Discoverer} \\
\hline
  03:23:01.53 & +56:26:00.8 &  289$\pm$6 & -279$\pm$6 & 13.65$\pm$0.02 & \citet{luhman14a}     \\
  04:54:25.04 & +40:04:10.4 &  388$\pm$4 & -177$\pm$4 & 14.44$\pm$0.02 & \citet{luhman14b}     \\
  05:32:52.18 & +41:43:39.6 &  226$\pm$4 & -274$\pm$4 & 11.66$\pm$0.02 & \citet{kirkpatrick14} \\
  06:48:37.95 & +07:37:02.1 &  -54$\pm$6 & -378$\pm$7 & 11.55$\pm$0.02 & \citet{luhman14a}     \\
  07:01:24.23 & -13:34:13.8 &  215$\pm$3 &  166$\pm$3 & 11.14$\pm$0.02 & \citet{luhman14a}     \\
  20:06:17.71 & +38:11:48.9 &  107$\pm$5 & -309$\pm$6 & 11.98$\pm$0.02 & \citet{kirkpatrick14} \\
  20:28:59.19 & +45:09:51.9 &  182$\pm$5 &  360$\pm$7 & 11.66$\pm$0.02 & \citet{luhman14a}     \\
  20:45:44.35 & +50:53:48.1 &  254$\pm$5 &  286$\pm$5 & 11.65$\pm$0.02 & \citet{luhman14a}     \\
  21:10:06.33 & +46:15:36.4 & -199$\pm$3 & -370$\pm$3 & 11.80$\pm$0.02 & \citet{luhman14a}     \\
  21:44:42.19 & +47:44:05.5 & -201$\pm$3 & -231$\pm$3 & 13.43$\pm$0.02 & \citet{luhman14a}     \\
  22:22:33.43 & +58:57:09.0 &  394$\pm$5 &   75$\pm$5 & 12.81$\pm$0.02 & \citet{luhman14a}     \\
  22:42:00.46 & +58:16:40.3 &   18$\pm$3 & -311$\pm$3 & 11.21$\pm$0.02 & \citet{kirkpatrick14} \\
\hline
\end{tabular}
\end{minipage}
\end{table*}

\end{document}